\documentclass[usenatbib]{mn2e}
\usepackage{graphicx,times}
\usepackage{bm}
\usepackage{epsf}
\usepackage{amsmath}
\usepackage{color}

\def\oh{\hat {\bf\Omega}}
\def\be{\begin{equation}}
\def\ee{\end{equation}}
\def\myC{{\cal C}}

\def\rY{Y}
\def\rZ{Z}
\def\rH{H}
\def\rE{E}
\def\rB{B}
\def\rS{S}

\def\cb{\textcolor{black}}

\def\beqa{\begin{eqnarray}}
\def\eeqa{\end{eqnarray}}
\def\ben{\begin{eqnarray}}
\def\een{\end{eqnarray}}
\def\nn{\nonumber}
\begin{document}
\onecolumn
\title[Reionization and CMB non-Gaussianity]
{Reionization and CMB non-Gaussianity}
\author[Munshi et al.]
{D. Munshi$^{1,2}$, P. S. Corasaniti$^{3}$, P. Coles$^{1,2}$, A. Heavens$^{4}$, S. Pandolfi$^{5}$ \\
$^{1}$ School of Mathematical and Physical Sciences, University of Sussex, Brighton BN1 9QH, U.K.\\
$^{2}$School of Physics and Astronomy, Cardiff University, Queen's
Buildings, 5 The Parade, Cardiff, CF24 3AA, UK \\
$^{3}$Laboratoire Univers et Th\' eories (LUTh), UMR 8102 CNRS, Observatoire de Paris,
Universit\' e Paris Diderot, 5 Place Jules Janssen, 92190 Meudon, France\\
$^{4}$Imperial Centre for Inference and Cosmology, Blackett Laboratory, Imperial College, Prince Consort Road, London SW7 2AZ, UK\\
$^{5}$Dark Cosmology Centre, Niels Bohr Institute, University of Copenhagen, Juliane Maries, Vej 30, 2100 Copenhagen, Denmark \\}
\maketitle 
\begin{abstract}
We show how cross-correlating a high redshift {\em external} tracer field, such as the 21cm neutral hydrogen
distribution and product maps involving Cosmic Microwave Background (CMB) temperature and polarisation fields, that
probe mixed bispectrum involving these fields,
can help to determine the reionization history of the Universe,
beyond what can be achieved from cross-spectrum analysis.
Taking clues from recent studies for the detection of primordial non-Gaussianity \citep{MuHe10}, 
we develop a set of estimators
that can study reionization using a power spectrum associated with the bispectrum (or skew-spectrum). We use
the matched filtering inherent in this method to investigate different reionization
histories. We check to what extent they can be used to rule out various models of reionization and study
cross contamination from different sources such as the lensing of the CMB. 
The estimators can be fine-tuned to optimize study of
a specific reionization history. We consider three different types of tracers in our study,
namely: proto-galaxies; 21cm maps of neutral hydrogen; quasars.  We also consider four alternative models of reionization.
\cb{We find that the {\em cumulative} signal-to-noise (S/N) for detection at $\ell_{max}=2000$ can reach $\mathcal{O}(70)$ 
for cosmic variance limited all-sky experiments. Combining 100GHz, 143GHz and 217GHz channels of the Planck experiment, we find 
that the (S/N) lies in the range $\mathcal{O}(5)-\mathcal{O}(35)$.}
The (S/N) depends on the specific choice of a tracer field, and multiple 
tracers can be effectively used to map out the entire reionization history with reasonable S/N. 
Contamination from weak lensing is investigated and found to be negligible, and the effects of Thomson scattering from patchy reionization
are also considered.   
\end{abstract}
\section{Introduction}
Pinning down the details and controlling physics of the cosmological reionization history remains one of 
the important goals of present-day cosmology. 
It is well known, thanks to a large set of astrophysical observables, that after
primordial recombination, which occurred at a redshift of $z \sim 1100$, the Universe
reionized at a redshift $z>6$. The epoch-of-reionization (EoR) is related to many fundamental questions 
in cosmology, such as the properties of the first galaxies, physics of (mini-)quasars, 
formation of very metal-poor stars and a slew of other important research topics in astrophysics.
Hence uncovering it will have far reaching implications on the study of structure formation in the early Universe \citep{LB01}. 
Observations of Lyman-$\alpha$ forests with high-resolution echelle spectrographs on large
telescopes (such as HIRES on Keck, and UVES on ESO's Very Large Telescope) are valuable for studying reionization at $z\approx 2.5-6.5$ \citep{Fan06,FCK06}. Redshifted 21cm observations
are also a very important probe of the EoR and several instruments
are either operational or in the construction phase. In the short term these consists of:
The Low Frequency Array (LOFAR)\footnote{http://www.lofar.org/}, 
the Murchison Widefield Array (MWA)\footnote{http://www.mwatelescope.org/}, Precision Array
to Probe Epoch of Reionization (PAPER) and Giant Metrewave Radio Telescope
(GMRT)\footnote{http://gmrt.ncra.tifr.res.in/}, while, on a somewhat longer time scale the Square Kilometre Array (SKA)\footnote{http://www.skatelescope.org/} will be operational.

In addition to Lyman-$\alpha$ and 21cm redshifted observations, CMB temperature and polarisation studies can also provide
valuable information regarding the EoR. The polarisation signal in CMB is generated due to the scattering of the local
CMB temperature quadrupole by the free-electron population. This signal peaks at angular scales corresponding to the
horizon at the rescattering surface (at a few tens of degrees) and the amplitude depends on
total optical depth. However, the large cosmic variance
associated with the signal means it is impossible to discriminate among 
various reionization histories using cross-correlation of the CMB temperature
and polarisation  \citep{Kap03,HZKK08,HuHo08}.
The total optical depth to reionization using WMAP
data is $\tau=0.08 \pm 0.013$  \citep{Bennett13}. Most current constraints from CMB data are analysed 
assuming a ``sudden'' and complete reionization at a redshift $z_r$ for WMAP\footnote{http://map.gsfc.nasa.gov/} this value of
$\tau$ will correspond to $z_r=11$. However as mentioned before, the precise 
details of the reionization process are not very well known and clearly the reionization
history of the universe at those redshifts could have easily been very different. 
The combination of temperature data and lensing reconstruction from the Planck data
gives an optical depth $\tau=0.089 \pm 0.032$ \citep{Planck13d}\footnote{http://www.rssd.esa.int/index.php?project=Planck}, consistent with WMAP9 estimates.
The polarization data from Planck
is expected to improve the accuracy of determination of $\tau$, however, it is 
important to keep in mind that CMB observations only provide integrated or projected 
information on reionization.  

The process of reionization is expected to be patchy and inhomogeneous in scenarios 
where reionization is caused by UV emission from the first luminous objects \citep[e.g.][]{Meerburg13}, and the resulting
fluctuations in visibility will generate extra anisotropy at arcminute scales.
Even in scenarios when reionization is caused by energy injection from decaying particles or X-ray emission,
inhomogeneities in electron density can cause fluctuations in visibility, but these fluctuations are too small to be detected 
in temperature and polarization power spectra or their cross-spectrum.
Nevertheless, as one can imagine, additional signals due to inhomogeneneites in the free-electron
density may imprint additional features at smaller angular scales \citep{Hu00,Santos}, and the mixed bispectrum 
with external data sets was advocated to extract redshift information \citep{Cooray04, Al06,AF07,T10,HIM07}.

In fact, what is required is the three-point correlation (or equivalently the bispectrum) involving temperature, polarisation and a tracer field for the free electron population can 
extract useful information on the ionization history of the Universe \citep{Cooray04}. 
For details of the generation of secondary non-Gaussianity due to reionization see \citep{KW10}.

The estimation of the bispectrum is a lot more complicated than the power spectrum due to the presence of additional degrees of freedom.
As has been pointed out in many recent works, the mode-by-mode estimation of the bispectrum, 
though very attractive, is seldom useful because of the associated low signal-to-noise \citep[e.g.][for a review]{Bart04}. Typically, this means, the entire information content 
of the bispectrum is often compressed into a single number which is used to distinguish various models 
of reionization. Though this has the advantage of increasing the (S/N),
it also degrades the information content of the bispectrum. 

A compromise solution was proposed recently by \cite{MuHe10}, who defined a power spectrum associated
with a specific bispectrum. This power spectrum represents the cross-spectra of the product
of two maps $[\rm X(\oh)Y(\oh)]$ against another map $\rm Z(\oh)$. It is a weighted sum of 
individual modes of the bispectrum keeping one of the index fixed while summing over
the other two indices. Such an estimator can also be designed to work with an experiential mask
for estimation in the presence of non-uniform noise and is relatively simple to
implement. In the literature such 
estimators are known as pseudo-$\myC_\ell$ (PCL) estimators. However, such estimators
are sub-optimal. With the recent attempts to detect primordial non-Gaussianity in the aftermath
and leading up to the Planck data release \citep{Planck13a,Planck13b}
there has been an increased activity in the area of optimising estimators which
can probe primordial non-Gaussianity \citep{Heav98,KSW,Crem06,Crem07b,SmZaDo00,SmZa06}.
Detection of secondary non-Gaussianity, can also benefit from using the \cite{MuHe10} estimators to 
take into account inhomogeneous noise and partial sky coverage in an optimal
way \citep{MuCoHeVa11}. We will discuss these issues and other related optimisation problems
in this paper for mixed data sets. Being able to probe the bispectrum in a scale-dependent way will provide a useful way to differentiate among different theories
of reionization. The matched filtering inherent in these estimators are likely
to be very useful in pinning down a specific reionization history. The formalism
also provides a natural set-up to study cross-contamination from effects of weak
lensing. The approach presented here has already been successfully implemented for Planck
data analysis which resulted in the detection of non-Gaussianity from the correlation of the Integrated Sachs-Wolfe (ISW) effect with gravitational lensing, and from residual point sources in the maps \citep{Planck13b}.  

We will define several set of different estimators. 
In addition to using the direct estimators, we will define, three dimensional constructs
that require the use of appropriate weight functions to cross-correlate and 
probe the secondary non-Gaussianity. We will also point to computationally 
extensive estimators which can take into account all possible complications. 
These generalisations involve a set of fully optimal estimators
which can work directly with harmonics of associated fields and 
carry out inverse covariance weighting using a direct brute force approach.
Clearly, though such a direct approach is completely optimal it is prohibitively expensive
to implement beyond a certain resolution. Nevertheless, for secondary bispectrum which
we consider here, it will be important to maintain the optimality to a high
resolution as most relevant information will be appear on small scales.
 
The results presented here can be seen as an extension of our earlier papers:  for example, \citep{MuHe10},
where we presented skew-spectrum for primary non-Gaussianity; \citep{MuCoHeVa11}, where results relevant to
secondary non-Gaussianity were obtained. Here, we include polarization data in addition to
the temperature maps and cross-correlate with external data sets in 3D to constrain
various scenarios of reionization. In recent years we have extended the concept of skew-spectra
to study topological properties of CMB maps \citep{MuCoHe13, MuSmCoReHaCo13} as well as for other cosmological data sets 
such as frequency-cleaned thermal Sunyaev Zeldovich $y$-maps \citep{MuSmJoCo12a} or weak lensing maps \citep{MuSmWaCo12b}.

This paper is organised as follows: In \textsection\ref{ana} we present many of
our analytical results and include the description of reionization models and 
the tracer fields that we study. We introduce our estimators in \textsection\ref{sec:estim}.
In \textsection\ref{sec:disc} we discuss our  
results and \textsection\ref{sec:conclu} is devoted to concluding remarks. In Appendix \ref{sec:patch}
we outline how a skew-spectrum estimator can be constructed using minimum variance 
estimated of fluctuations in optical depth. In Appendix \ref{sec:lensing_recon} we provide
equivalent estimators for the reconstruction of the lensing potential.

\section{Notations and Analytical Results}
\subsection{Mixed Bispectrum}
\label{ana} 
Given a bispectrum involving three different fields $\rm X(\oh)$, $\rm Y(\oh)$ and $\rm Z(\oh)$, we can define
a mixed bispectrum $B^{\rm XYZ}_{\ell_1\ell_2\ell_3}$ which encodes non-Gaussianity at the three-point level
(see \cite{MuHe10} for more discussion regarding definitions related to the bispectrum
and its estimation).
\begin{equation}
\langle {\rm X}_{\ell_1m_1}{\rm Y}_{\ell_2m_2}{\rm Z}_{\ell_3m_3} \rangle_c = \sum_{m_1m_2m_3}B_{\ell_1\ell_2\ell_3}^{\rm XYZ}\left ( \begin{array}{ c c c }
     \ell_1 & \ell_2 & \ell_3 \\
     m_1 & m_2 & m_3
  \end{array} \right); \qquad {\rm X,\rY,\rZ} = {(\Theta, E, S)}~ {\rm or} ~{(\Theta,\rB, S)}.
\end{equation}
The matrices here represent 3j symbols \citep{Ed68} and reflect the rotational invariance of the three-point correlation function. We will specialize our discussion later
to the case of cross-correlating temperature $\Theta$, polarisation fields ($E\pm iB$)  with a {\it tracer} field $S$
that traces the fluctuations in the free electron density. 
The secondary polarisation is generated by rescattering of CMB photons at a much lower
redshift than decoupling. 

The polarisation field is
\begin{equation}
P_{\pm}(\oh) = (q \pm iu)(\oh) = 
{\sqrt{24\pi} \over 10} \int dr g(r) \sum_{m=-2}^{2} \delta \Theta_{2m}({\bf x}){}_{\pm2}Y_{2m}(\oh);\\
\quad\quad \delta \Theta_{2m} = -{1\over 4 \pi}\int d\oh Y_{2m} \delta \Theta_{2m}({\bf x},\oh);
\end{equation}
with
\be
g(r) \equiv \dot\tau(r) \exp[{-\tau(r)}]=  x_{\rm e}(z){\rH}_0 \tau_{\rm H} (1+z)^2 \exp({-\tau}); \qquad 
\tau(r) = \int_0^r dr' \;\dot\tau(r'); \qquad \tau_{\rm H} =0.0691(1-Y_{\rm p}) \Omega_bh.
\ee
Here $g(r)$ is the visibility function which represents the probability of an electron being scattered within
a distance $dr$ of $r$; $\tau(r)$ is the optical depth out to distance $r$ with $\tau_H$ denoting 
the optical depth to the Hubble distance today due to Thomson scattering, which assumes full hydrogen ionization and a primordial
helium fraction $Y_{\rm p} = 0.24$; $x_{\rm e}(z)$ is the ionization fraction as a function of 
redshift $z$. The conformal distance $r$ at redshift $z$ is given in terms of the Hubble parameter as $r(z) = \int_0^z {dz' / \rH(z')}$ with $\rH^2(z) = H_0^2[\Omega_{\rm M}(1+z)^3 + \Omega_{\rm K}(1+z)^2+\Omega_\Lambda]$, parametrised in terms of the total cosmic matter density (cold dark matter + baryons) $\Omega_{\rm M}=\Omega_c +\Omega_b$, the cosmological constant density $\Omega_{\rm \Lambda}$ and the curvature $\Omega_{\rm K} = (1-\Omega_{\rm M}-\Omega_\Lambda)$ in units of the critical density $3{\rm H}_0^2/8\pi\,G$. Here $\rH_0^{-1}= 2997.9h^{-1}$Mpc is the inverse Hubble distance with $h=\rH_0/100$. In the following we will assume a standard flat LCDM cosmological model with $\Omega_c = 0.30$, $\Omega_b = 0.05$, $\Omega_\Lambda = 0.65$ and $h=0.65$ respectively.

The primary effect of reionization is manifested by the suppression of the temperature power-spectrum $\myC^{\Theta\Theta}_{\ell}$
by a factor $\exp(-2\tau)$ and enhancement of $\myC^{\rm EE}_{\ell}$ power spectrum at small $\ell$
which scales as $\tau^2$. Most CMB calculations adopt an abrupt reionization.
However low redshift studies involving Lyman-$\alpha$ optical depth 
related Gunn-Peterson troughs of the $z\sim6$ quasars indicate a more complex reionization
history; moreover the reionization can be patchy or inhomogeneous \citep{BL01}.

The mixed bispectrum $B_{\ell_1\ell_2\ell}^{\rS \Theta \rE}$ can be written as:
\begin{equation}
B_{\ell_1\ell_2\ell}^{E \Theta S} = \sqrt{\Sigma_{\ell_1}\Sigma_{\ell_2}\Sigma_\ell \over 4 \pi}
\left ( \begin{array}{ c c c }
     \ell_1 & \ell_2 & \ell \\
     0 & 0 & 0
  \end{array} \right) b^E_{\ell_2\ell_1};  \quad\quad \Sigma_\ell \equiv (2\ell+1),
\end{equation}
where $b^E_{\ell_2\ell_1}$ is the reduced bispectrum, using the Limber's approximation the latter reads as \citep[see][for derivation and detailed discussion]{Cooray04}\footnote{A similar result holds for the case related to B-type polarisation i.e. 
$B_{\ell_1\ell_2\ell}^{B \Theta S}$ in terms
of $b^B_{\ell_2\ell_1}$ which is obtained from Eq.(\ref{eq:def_bi}) by replacing $\epsilon_\ell^E(kr)$
with $\epsilon_\ell^B(kr)=2j_\ell'(x) /x^2 + 4j_\ell(x)/x$.}:
\ben
&& b^E_{\ell_2\ell_3} = {2 \over 9\pi} \int dr \; g(r) \; {G^2(r) \over d_A^2(r)} W_S(r) P_{gS} 
\left ( k={\ell_3 \over d_A} \right ) {\cal I}_{l_2}^E(r); \qquad
{\cal I}_\ell^E(r) = \int k^2 dk P_{\Phi\Phi}(k,r_0)j_\ell(kr_0)j_2(kr_s)\epsilon^E_\ell(kr); \quad\quad\\
&& r_s\equiv r-r_0; \quad\quad r_0=r\,(z=1100);
\label{eq:def_bi}
\een
where $\epsilon_\ell^E(kr)$ can be written in terms of the spherical Bessel function $j_\ell(x)$ and its derivatives $j'_\ell(x)$
and $j''_\ell(x)$:
\be
\epsilon_\ell^E(x) = -j_\ell(x) + j_\ell''(x) + 2j_\ell(x) /x^2 + 4j_\ell'(x)/x.
\label{eq_epsilon_EB}
\ee
In the above expression $d_A(r)= \rH_0^{-1}\Omega_{\rm K} \sinh[\rH_0\Omega_{\rm K}^{1/2}r]$ is the angular diameter distance which in a flat universe ($\Omega_K\rightarrow 0$) reduces to $d_A(r) \rightarrow r$. $W_S(r)$ represents the spatial distribution of the tracers and $G(r)$ is the linear growth factor defined such that the Fourier transform of the overdensity field grows as $\delta({\bf k},r)=G(r)\delta({\bf k},0)$ given by:
\be
G(r) = {\rH(z) \over \rH_0} \int_{z(r)}^{\infty} dz' (1+z') \left [\rH(z') \right ]^{-3}\Big /
 \int_{0}^{\infty} dz'' (1+z'') \left [ \rH(z'') \right ]^{-3},
\ee
which we compute using a standard numerical integration algorithm.

Throughout we will be using the normalisation $\langle \Phi({\bf k}) \Phi({\bf k'}) \rangle = 
(2\pi)^3 \delta^D( {\bf k} + {\bf k'}) P_{\Phi\Phi}(k)$ for the power spectrum
$P_{\Phi\Phi}(k)$ of the primordial potential perturbation $\Phi$ and
$\langle \delta_g({\bf k}) \delta_S({\bf k'}) \rangle = (2\pi)^3 \delta^D( {\bf k} + {\bf k'}) P_{gS}(k)$,
where $P_{gS}(k,z)$ is the cross-spectrum at a redshift $z$ between fluctuations in the scattering visibility function and the tracer field. In the following we assume a halo model such that $P_{gS}(k,z) = b_g(z)b_S(z)G^2(z)P^{\rm L}_{\delta\delta}(k)$ where 
$b_g(z)$ and $b_S(z)$ are the biases at large scales of the underlying fields and $P^{\rm L}_{\delta\delta}(k)$ is 
the linear matter power spectrum given by:
\be
P_{\delta\delta}^{\rm L}(k) = 2\pi^2 A_s \left(\frac{k}{k_p}\right)^{n_s-1} k\,T^2(k),
\ee
where $A_s$ is the scalar amplitude, $n_s$ is the scalar spectral index, $k_p=0.05$ Mpc$^{-1}$ the pivot scale and $T(k)$ is the CDM transfer function which we compute from \citep{EH98}. We assume $n_{\rm s}= 0.9635$ and $A_{\rm s}= 2.19\times10^{-19}$.
We will use these results to construct estimators based on the PCL approach or near optimal estimators based on 
generalisation of \cite{MuHe10}.

\subsection{Tracer Distribution Models}
\begin{figure}
\begin{center}
{\epsfxsize=10 cm \epsfysize=5 cm {\epsfbox[25 435 589 716]{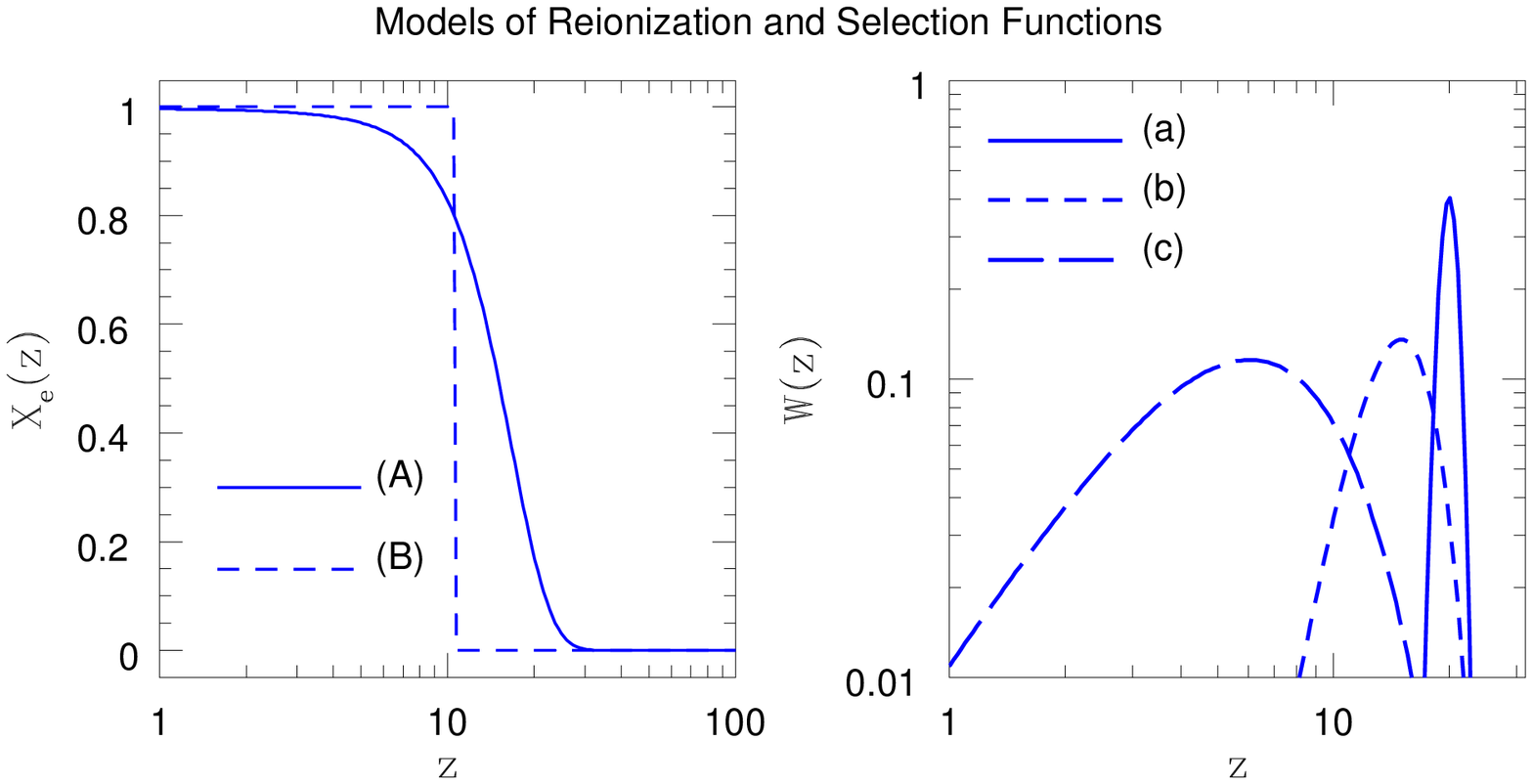}}}
\caption{Left panel: ionization fraction $x_{\rm e}(z)$ for {\em slow} or {\em smooth} (A) and {\em sudden} (B) reionization models (see text). Right panel: redshift distribution of sources $W(z)$ for three different populations of tracers corresponding to proto-galaxies (a), 21cm-emitters (b) and quasar-like sources (c).}
\label{fig:models}
{\epsfxsize=10 cm \epsfysize=5 cm {\epsfbox[25 435 589 716]{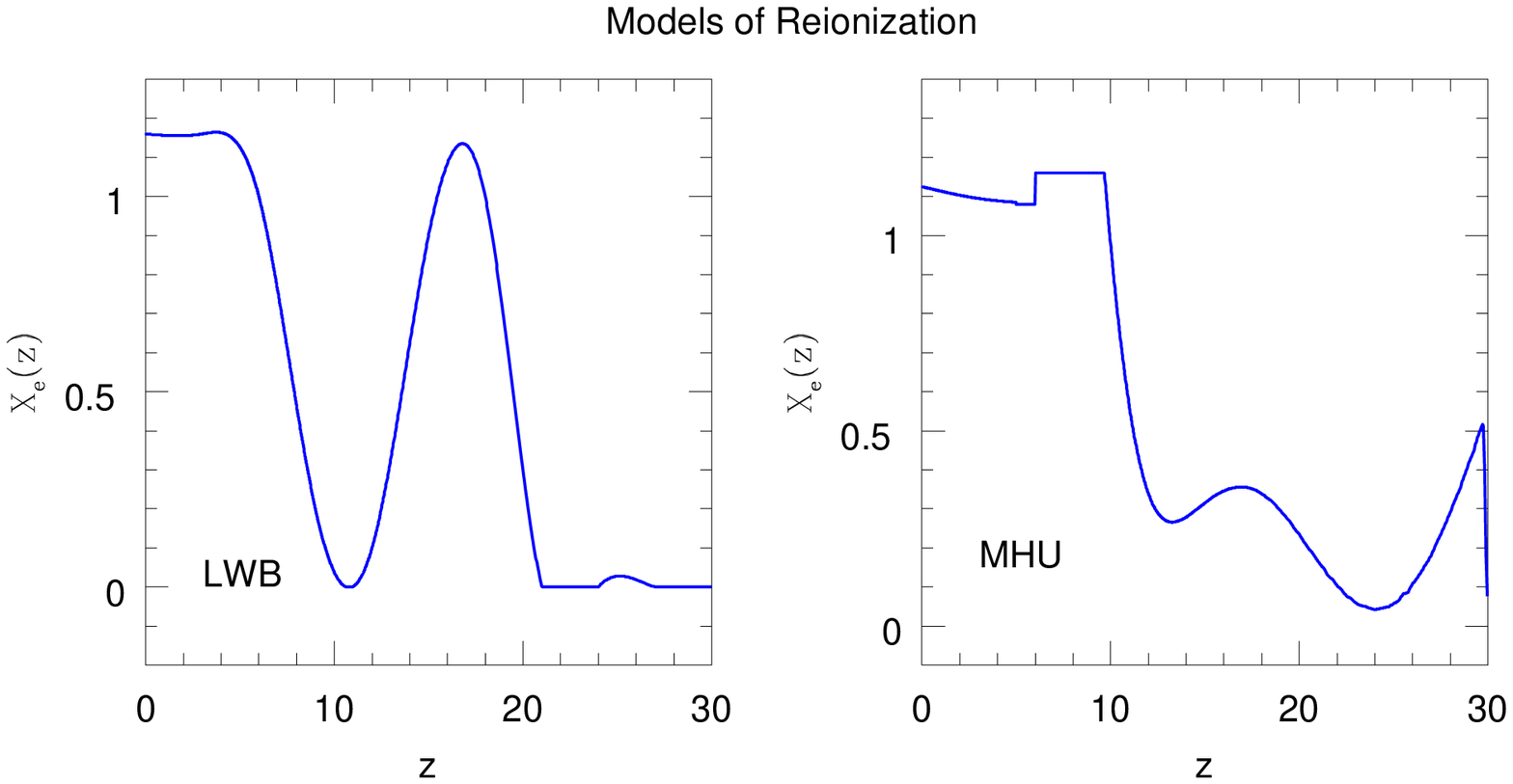}}}
\caption{Ionization fraction $x_{\rm e}(z)$ for two different non-standard reionization models denoted as LWB \citep{LWB06} (left panel) and MHU \citep{MoHu08} (right panel). In the case of LWB a binned ionization history was considered, while a
Principle Component Analysis (PCA) was employed in MHU.}
\label{fig:models2}
\end{center}
\end{figure}

We consider three different tracer sources corresponding to a population of proto-galaxies contributing to the IR background (a), 21cm-like tracers (b) and a quasar-like sources (c). In the case of tracer model (a) and (b) we assume a Gaussian redshift distribution: 
\ben
&& { \rm Model\;(a)\;and\; (b)}: W(z) = {1 \over \sqrt{2\pi \sigma_z^2}}\exp \left [-{1 \over 2}{(z - \bar z)^2 \over \sigma_z^2 } \right ].
\label{eq:wz1}
\een
with parameters $\bar z = 15$ and $\sigma_z = 3$ for (a) and $\bar z=20$ and $\sigma_z=1$ for (b). In the case of quasar-like sources (c) we assume a broad redshift distribution 
\ben
&& {\rm Model\;(c)}: W(z) = \left ({z \over {\bar z}} \right )^\alpha\left ({\beta \over {\bar z}}\right )\exp\left [-\left ({z \over {\bar z}}\right )^\beta\right ].
\label{eq:wz2} 
\een
with parameters  $\bar z=3$, $\alpha=2$ and $\beta=1.5$.
The normalised redshift distributions of the tracer fields are plotted in Fig.~\ref{fig:models} (right-panel).

\subsection{Reionization Models}
In order to test the potential of testing the reionization history through mixed bispectrum analyses we focus on different reionization history models which are indistinguishable from one another using CMB temperature, polarization and their cross-correlation spectra. These models are characterized by differention redshift dependencies of the ionization fraction:
\ben
&&\quad{\bf{\rm \underline{Model \; (A)}}} \;\; ({\rm or\; the\; smooth \; model}): x_{\rm e}(z)=1-{1 \over 2}{\rm Erfc}\left[{(z_r-z)\over \sigma_z}\right ];\quad\quad z_r=15,\quad \sigma_z=7.5 \label{eq:xe1};
\een
\ben
&&\quad{\bf{\rm \underline{Model \; (B)}}} \;\; ({\rm or\; the\; sharp \; model}): 
x_{\rm e}(z)=
\begin{cases}
1 &\text{if $z<z_r$}\\ 
0 &\text{otherwise}\end{cases};\quad\quad z_r=11.35
\label{eq:xe2}
\een

\enspace\enspace\enspace\enspace\thinspace\quad{{\rm \underline{Model (LWB)}}: this is double reionization scenario studied in \cite{LWB06} which we implement by considering 12 redshift bins centered at $z={0,3,6,9,12,15,18,21,24,27,30,32}$ with values of $x_{\rm e}={1.16,1.16,1.0,0.2,0.1,0.9,1.0,0.002,0.002,0.002,0.002,0.002}$ respectively which we use to built a cubic spline interpolation of $x_{e}(z)$. The values of $x_{e}$ in the first two redshift bins take into account the effect of second helium reionization, while the third one only has the contribution from first hydrogen reionization. The total optical depth for this model is 0.14. For $z>18$ we set $x_{e}=2\times 10^{-4}$ which is the value of $x_{e}$ expected before reionization (following primordial recombination).}

\enspace\enspace\enspace\enspace\thinspace\quad{{\rm \underline{Model (MHU)}}: this reionization scenario has been studied in \citep{MoHu08} and is based on a parametrized reionization history built by decomposing $x_e(z)$ into its principal components $x_e(z)=x_e^{\rm fid}(z)+\sum_{\mu}m_{\mu}S_{\mu}(z)$, where the principal components, $S_{\mu}(z)$, are the eigenfunctions of the Fisher matrix that describes the dependence of the  $C_\ell^{\rm EE}$ on $x_e(z)$, $m_{\mu}$ are the amplitudes of the principal components for a given reionization history, and $x_e^{\rm fid}(z)$ is the fiducial model at which the Fisher matrix is computed. For the MHU model we have used the first five principal components for the reconstruction of the reionization history \citep[for more details see][]{PandolfiA}. The values of amplitudes are consistent with the one-sigma confidence level values around the best fit obtained using Planck data \citep{Planck13c}. The total optical depth for this model is $\tau=0.15$.}

The smooth model (A) corresponds to reionization by UV light from star forming regions within collapsed halos \citep{Cooray04}, while the instantaneous transition model (B) to a fully reionised Universe from a neutral one could arise e.g. in the presence of X-ray background \citep{Oh01} or from decaying particles \citep{HH03, CK04, KKS04}. The redshift evolution of the ionization fraction for models (A) and (B) is shown in Fig.~\ref{fig:models} (left-panel), while in Fig.~\ref{fig:models2} we plot
the ionization fractions of LWB (left-panel) and MHU (right-panel). 

\begin{figure}
\begin{center}
{\epsfxsize=15 cm \epsfysize=5 cm {\epsfbox[27 521 584 709]{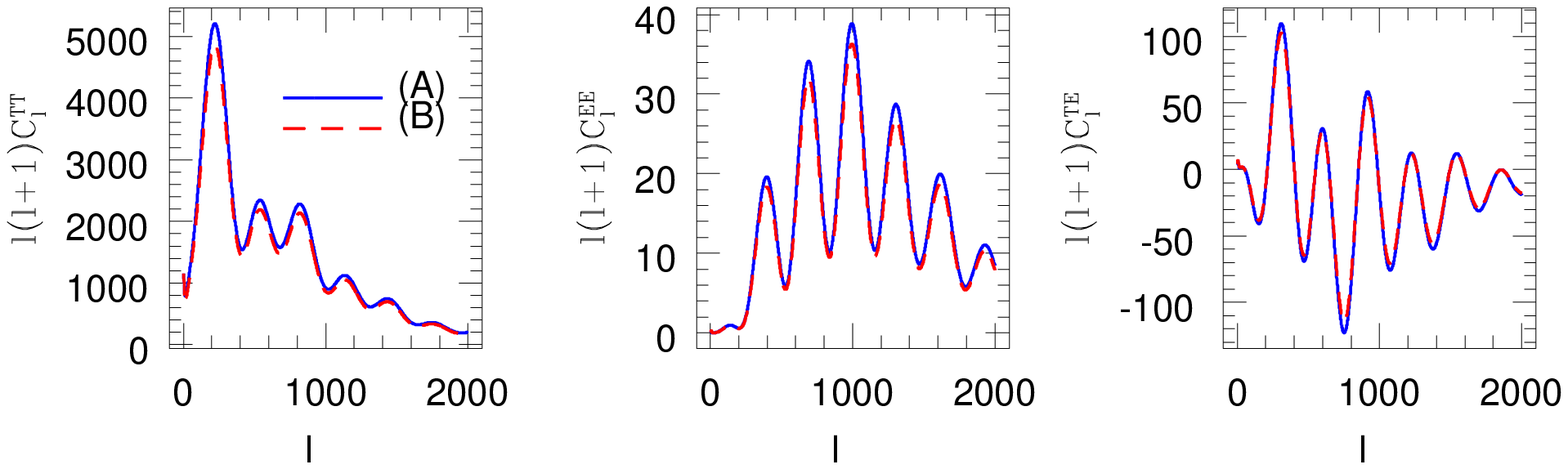}}}
\caption{The CMB temperature power spectrum $\myC^{\rm TT}_{\ell}$ (left panel), {\em Electric} or E-type polarization power-spectra 
$\myC^{\rm EE}_{\ell}$ (middle panel) and temperature-polarization cross-spectrum $\myC^{\rm T \rm E}_{\ell}$ 
(right panel) for models (A) ``slow'' and  (B) ``sudden''.}
\label{fig:te}
{\epsfxsize=15 cm \epsfysize=5 cm {\epsfbox[27 519 584 709]{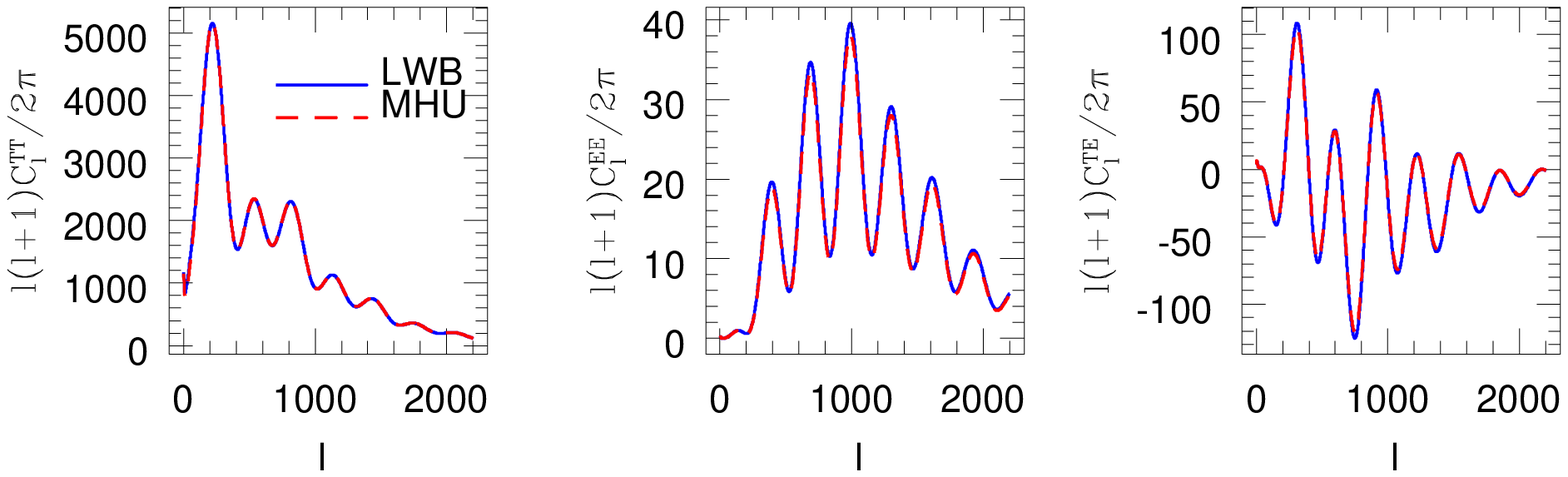}}}
\caption{As in Fig.~\ref{fig:te} for LWB and MHU models respectively.}
\label{fig:te2}
\end{center}
\end{figure}

In Fig.~\ref{fig:te} and \ref{fig:te2} we plot the temperature $\myC^{\rm TT}_{\ell}$ (left panel), E-polarization $\myC^{\rm EE}_{\ell}$ (middle panel) and cross temperature-polarization spectra $\myC^{\rm T \rm E}_{\ell}$ (right panel) for the different reionization history models. We can see that despite the different redshift dependence of the ionization fraction these models are indistinguishable using only CMB measurements. 

\begin{figure}
\begin{minipage}{17cm}
\begin{center}
{\epsfxsize=13 cm \epsfysize=4.7 cm {\epsfbox[27 520 584 709]{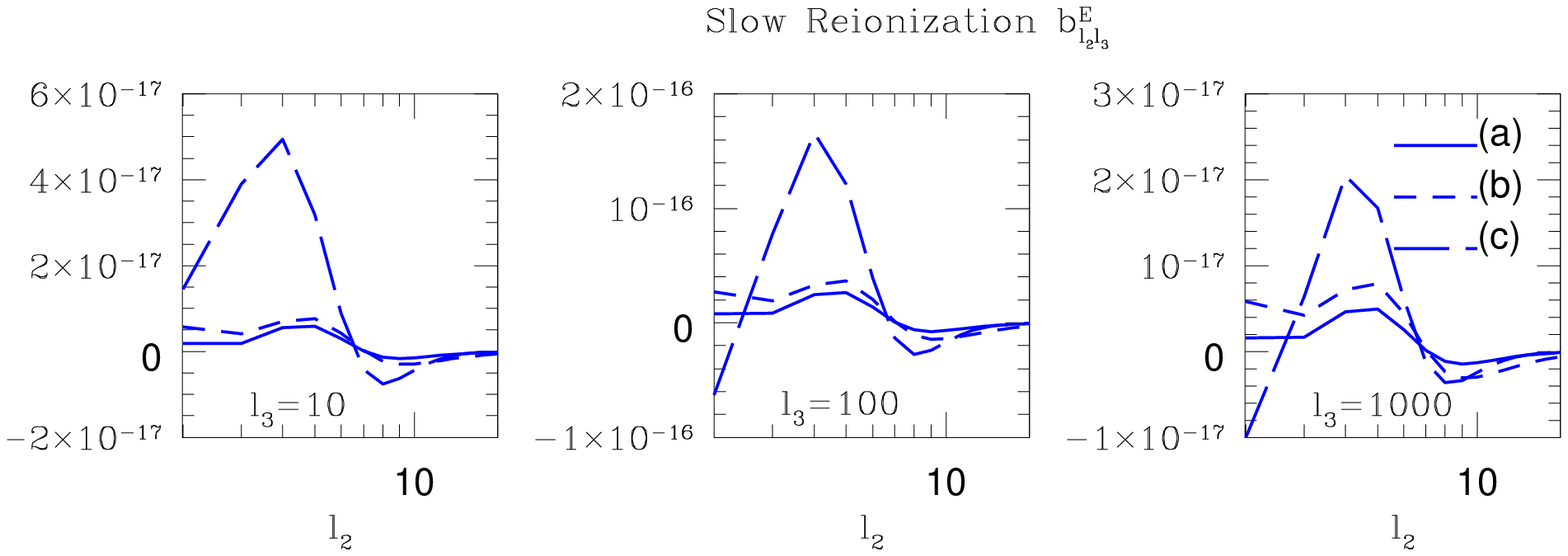}}}
{\epsfxsize=13 cm \epsfysize=4.8 cm {\epsfbox[20 510 584 719]{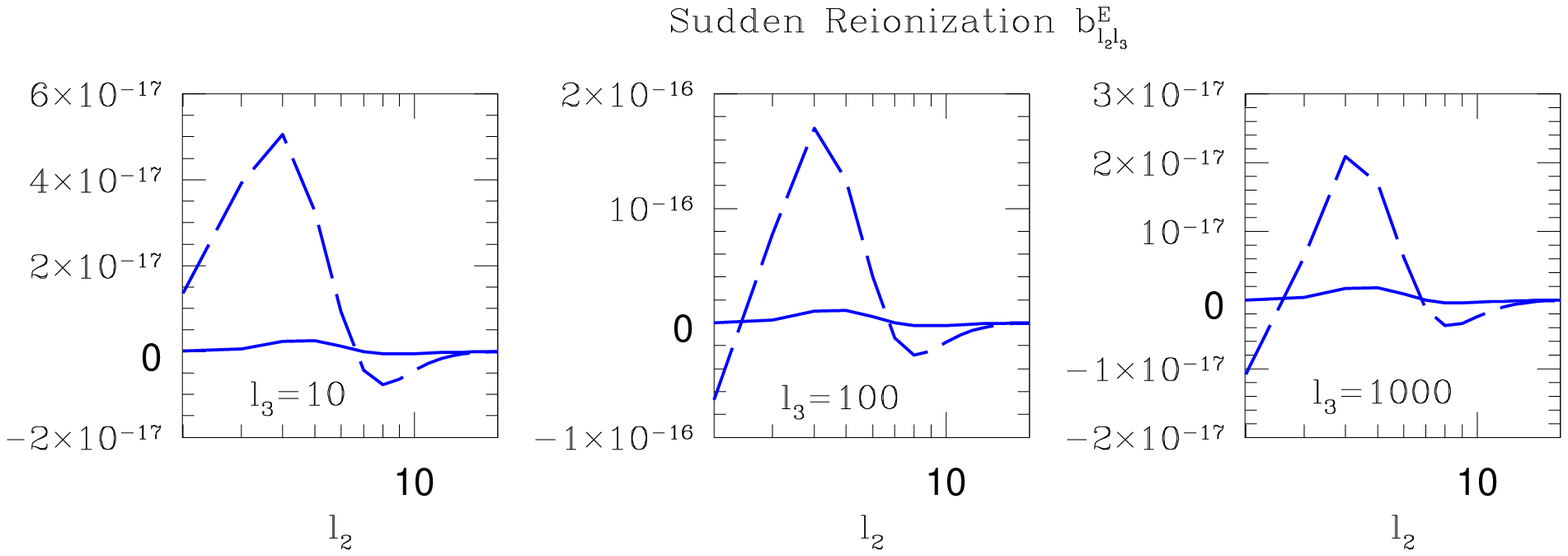}}}
{\epsfxsize=13 cm \epsfysize=4. cm {\epsfbox[27 535 584 729]{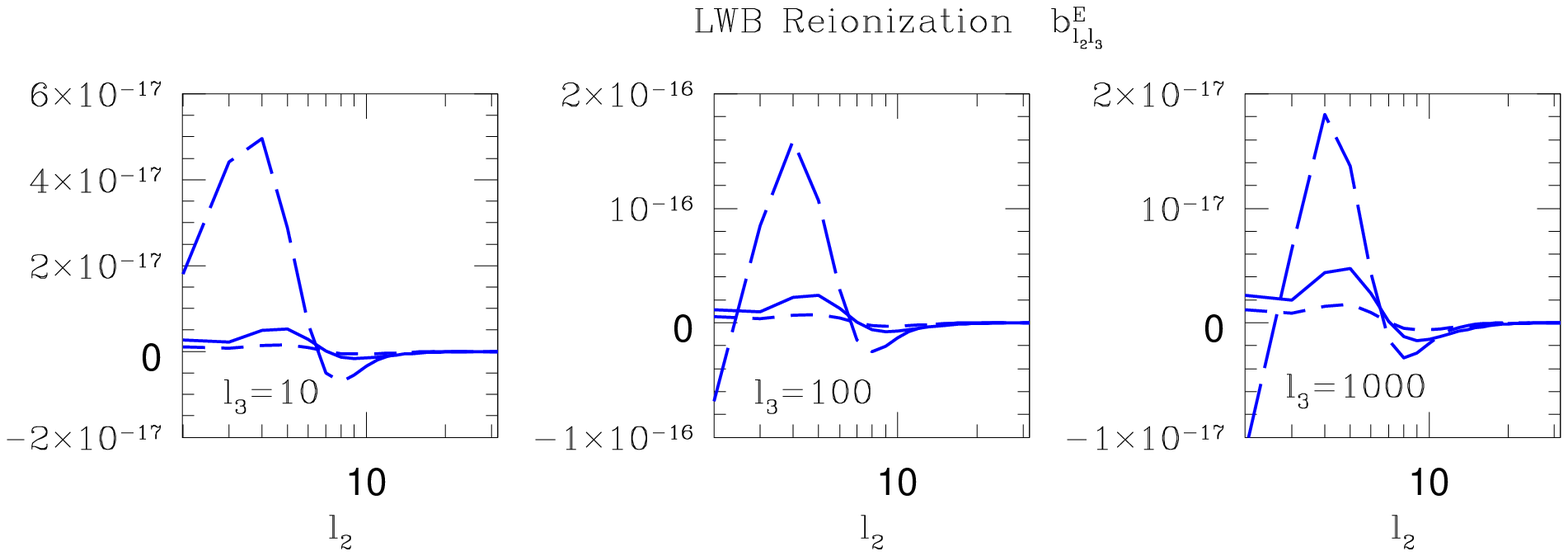}}}
{\epsfxsize=13 cm \epsfysize=5 cm {\epsfbox[27 515 590 719]{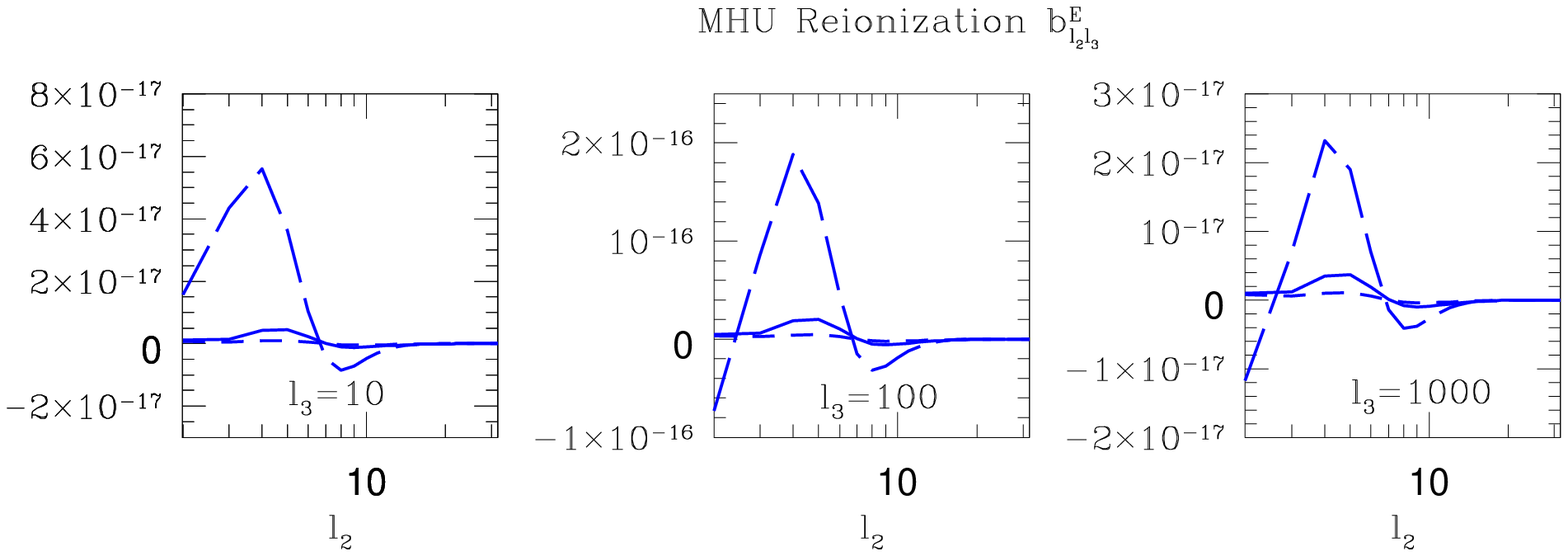}}}
\caption{Reduced bispectrum as function of $\ell_2$ at constant $\ell_3=10,100,1000$ for different reionization history models.}
\label{fig:bl2_l3fixed}
\end{center}
\end{minipage}
\end{figure}

\begin{figure}
\begin{minipage}{17cm}
\begin{center}
{\epsfxsize=13 cm \epsfysize=4.5 cm {\epsfbox[27 520 584 709]{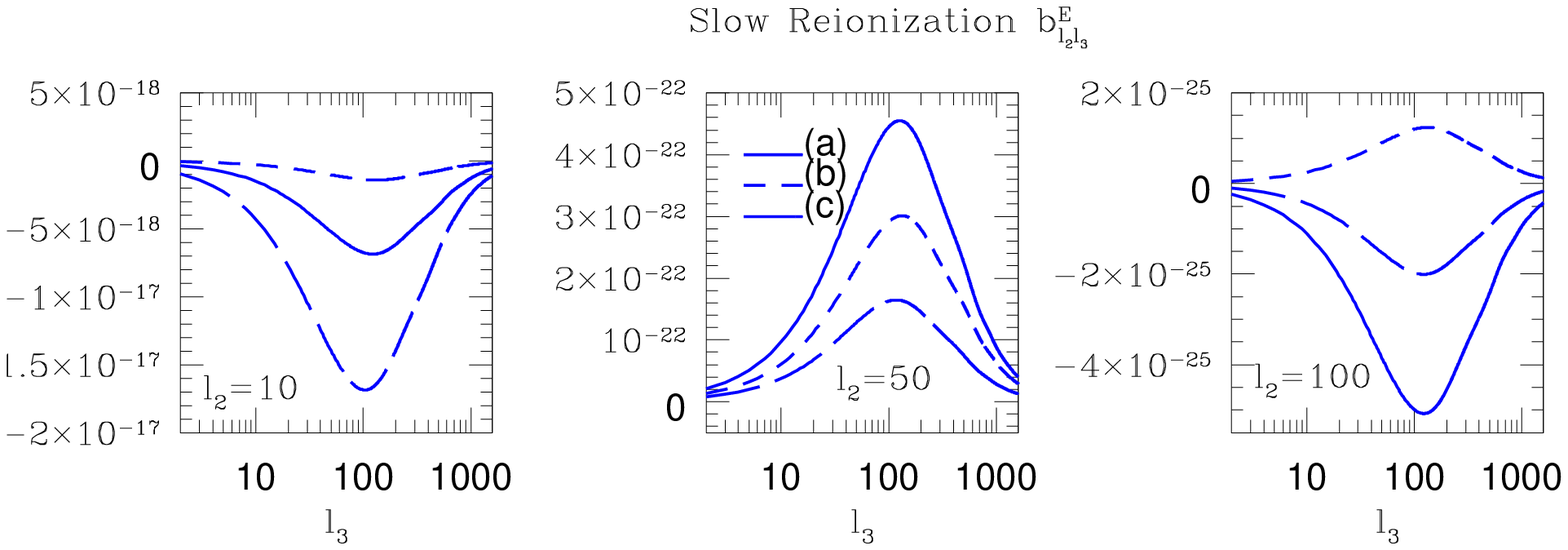}}}
{\epsfxsize=13 cm \epsfysize=4.5 cm {\epsfbox[27 520 584 719]{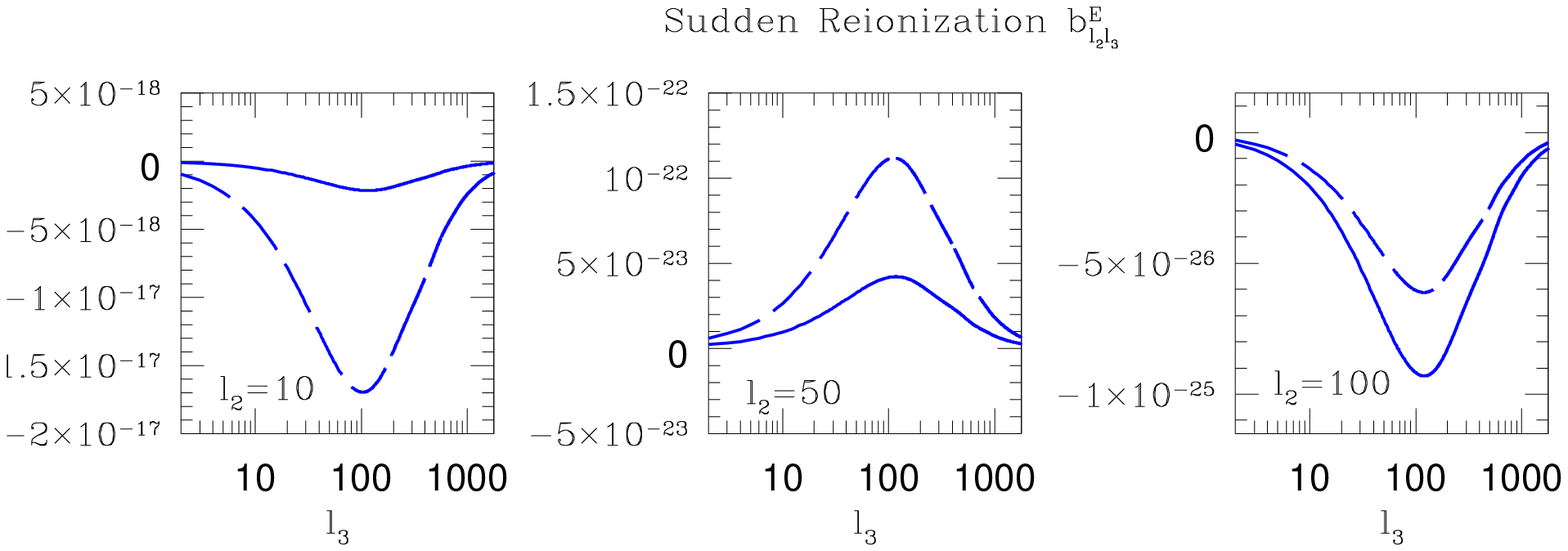}}}
{\epsfxsize=13 cm \epsfysize=4.3 cm {\epsfbox[27 535 584 739]{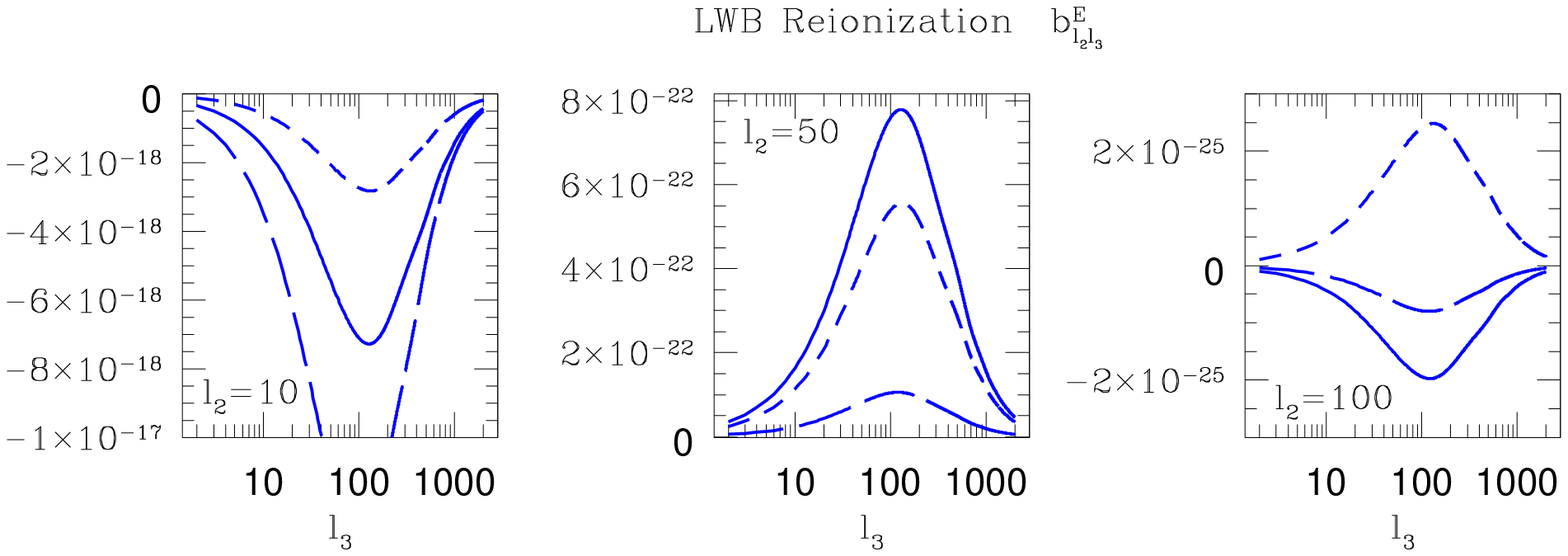}}}
{\epsfxsize=13 cm \epsfysize=5 cm {\epsfbox[27 515 590 719]{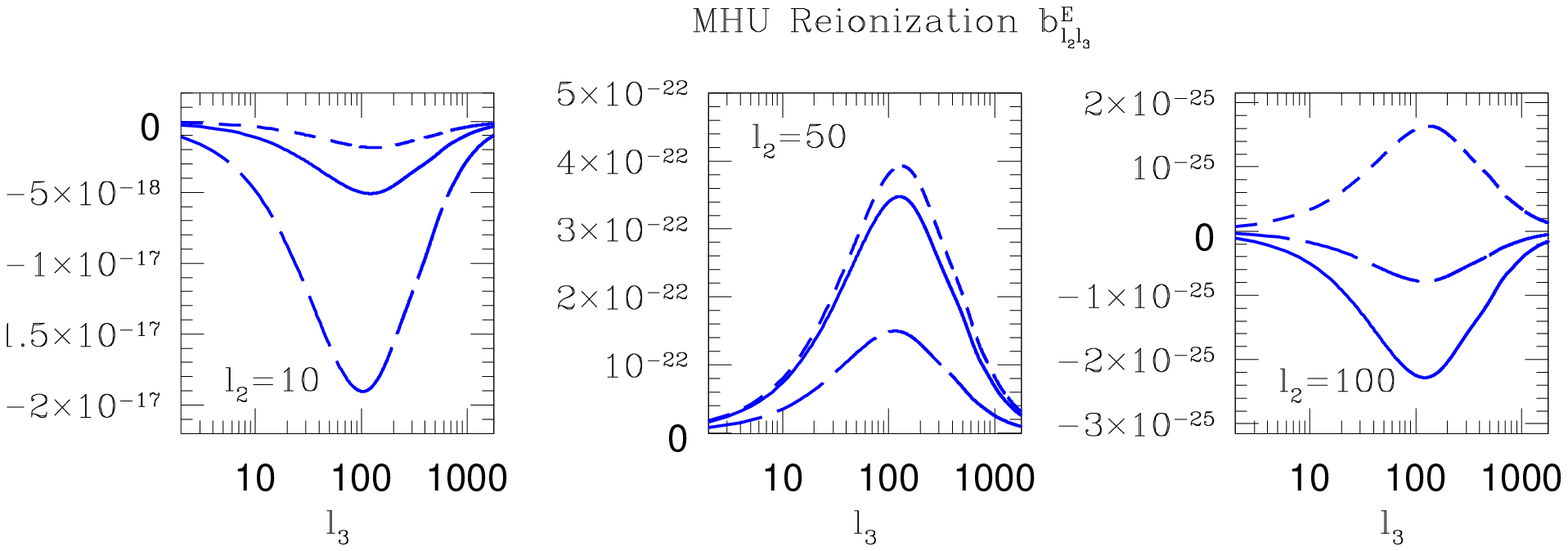}}}
\caption{Reduced bispectrum as function of $\ell_3$ at constant $\ell_2=10,50,100$ for different reionization history models.}
\label{fig:bl3_l2fixed}
\end{center}
\end{minipage}
\end{figure}

We compute the reduced bispectrum for the different reionization history models using Eq.~(\ref{eq:def_bi}), for simplicity we set the bias functions $b_g(z)=b_S(z)=1$. In Fig.~\ref{fig:bl2_l3fixed} we plot $b_{\ell_2\ell_3}$ as function $\ell_2$ at constant $\ell_3=10,100,100$, while Fig.~\ref{fig:bl3_l2fixed} we plot $b_{\ell_2\ell_3}$ as function $\ell_2$ at constant $\ell_3=10,100,100$. We can see that differently from the CMB spectra the different models gives different predictions for the amplitude of the reduced bispectrum. Furthermore they show that information is encoded in the multipole structure of $b_{\ell_2\ell_3}$. As one may expected the overall amplitude of the signal depends on how well the tracers overalp with a non-vanishing value of the ionization fraction. In other words the larger the redshift interval where the convolution of the tracer distribution with the visibility function is non-vanishing and the larger is the signal. As an example let us look at the reduced bispectrum at constant $\ell_3$ values for the ``slow'' and ``sudden'' reionization models shown in two top-panels of Fig.~\ref{fig:bl2_l3fixed}. In the ``sudden'' case the visibility function vanishes at $z>11.35$, this implies that fluctuations of high-redshift tracer fields such as model (a) will be less correlated with fluctuations of the visibility function and the large angular scale polarization pattern of the CMB. In contrast, in the ``slow'' model the visibility function is still non-vanishing at high-redshift thus leading to a larger value of the reduced bispectrum. The multipole structure of the reduced bispectrum at constant $\ell_2$ values depends on the specific reionization model and the tracer field and it is harder to disentangle since other factors such as the projected matter power spectrum, the amplitude of the tracer distribution (i.e. the abundance of sources), the specific redshift evolution of the visibility function comes into play. Nevertheless, it is worth noticing that the amplitude changes between positive and negative values at different values of $\ell_2$. This is because the integral ${\cal I}_{l_2}^E(r)$ in Eq.~(\ref{eq:def_bi}) oscillates as function of $\ell_2$ and as we can see in Fig. \ref{fig:bl3_l2fixed} the change of sign depends on the reionization model. Overall, this suggests that the bispectrum of fields which correlates with reionization history carries a distinctive imprint of this process. In the next sections we will assess the detectability of such signal for Planck-like experiments using skew-spectra.
 
\section{Estimators}
\label{sec:estim}
In this section we will introduce the estimators for our skew-spectra.
We first discuss the direct PCL estimator which is sub-optimal. Next we introduce the
near-optimal skew-spectra that involves constructing 3D fields with appropriate weights.
We will also discuss contamination from CMB lensing. The estimators presented here
will be useful in probing patchy reionization scenarios.

\subsection{Direct or Pseudo-$\myC_{\ell}$ Approach}
\begin{figure}
\begin{center}
{\epsfxsize=7.5 cm \epsfysize=7.5 cm {\epsfbox[27 150 584 709]{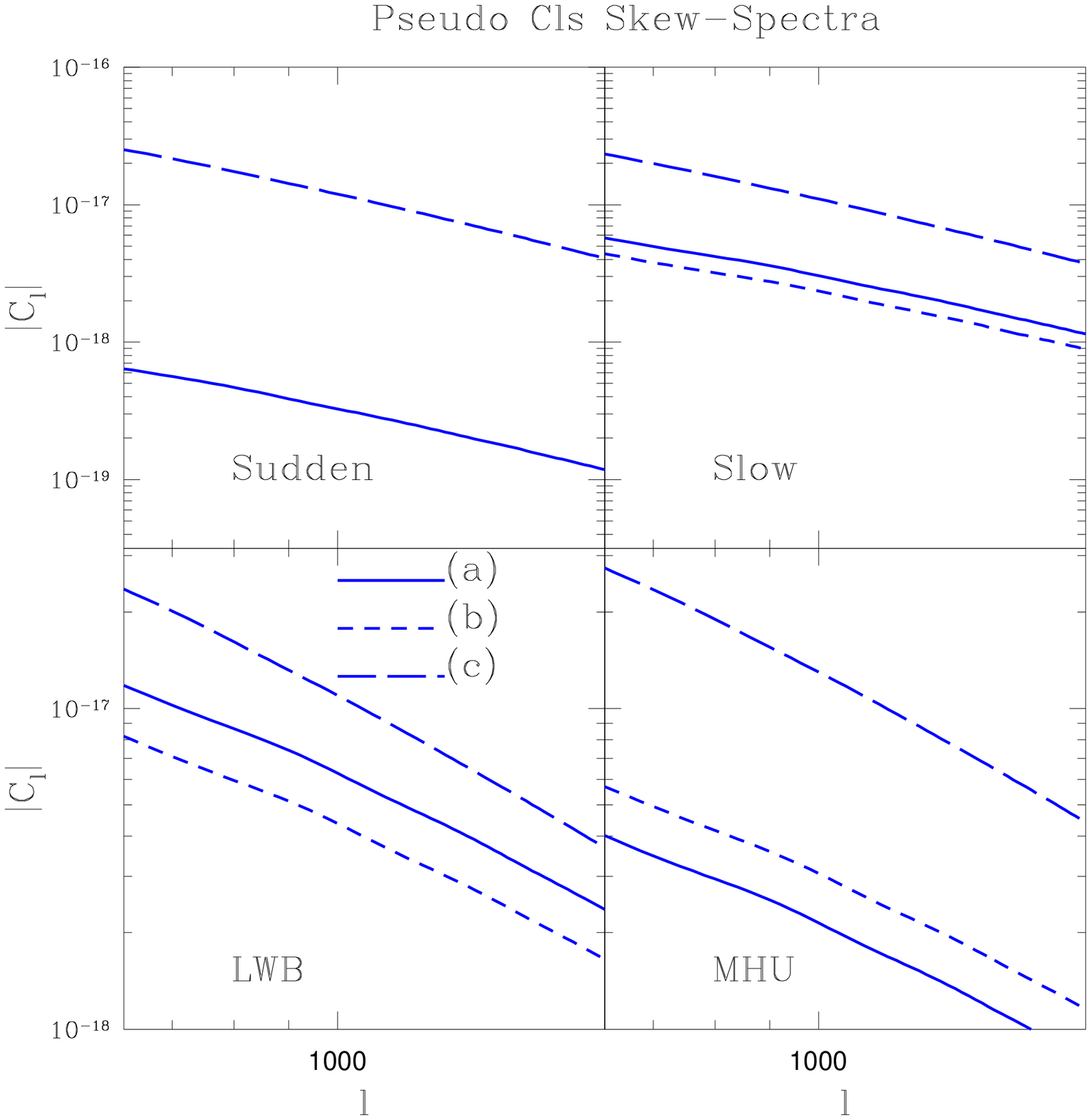}}}
\end{center}
\caption{The pseudo-${\cal C}_{\ell}$ (PCL) estimator
$\myC^{\Theta\rm E, X}_{\ell}$ defined in Eq.~(\ref{eq:direct}) for various models of reionization: ``slow'' (upper-right panel), 
``sudden'' (upper-left panel), LWB (lower-left panel) and MHU (lower-right panel). For each model we plot the PCL for the tracer
model (a) (solid line), model (b) (short-dashed line) and model (c) (long-dashed line) respectively.}
\label{fig:pcl}
\end{figure}

In principle it is useful to study the bispectrum for every possible triplet
of $(\ell_1,\ell_2,\ell_3)$ represented by triangular configuration in the harmonic
space. However, this is a challenging task due to the low signal-to-noise
associated with individual modes. The usual practice is to sum all possible
configurations and study the resulting skewness. However, this method of
data compression is extreme and a trade-off can be reached by summing
over two specific indices while keeping the other fixed. In real space 
this is equivalent to cross-correlating the product field $[{\rm X}(\oh){\rm Y}(\oh)]$ against ${\rm Z}(\oh)$.
The resulting power spectrum (called the skew spectrum) can be studied as a function of $\ell$.
The usual skewness then can be expressed as a weighted sum of this skew-spectrum.
In a recent work, \cite{MuHe10} proposed a power-spectrum associated with a given 
bispectrum. This encodes more information compared to the
one-point skewness often used in the literature. A PCL approach or
its variants in real or harmonic space has been used in many areas,
for analysing auto or mixed bispectrum from diverse dataset \citep{MuCoHeVa11}:
\begin{equation}
\myC_\ell^{\rm XY,Z} = \sum_{\ell_1\ell_2} B_{\ell\ell_1\ell_2}^{\rm XYZ} \sqrt{\Sigma_{\ell_1}\Sigma_{\ell_2}\over \Sigma_{\ell}}
\left ( \begin{array}{ c c c }
     \ell & \ell_1 & \ell_2 \\
     0 & 0 & 0
  \end{array} \right); \qquad {\rm X, Y, Z} = (\Theta,{\bf E},{\bf S}) \; or \; (\Theta,{\bf B},{\bf S}).
\label{eq:direct}
\end{equation}
The different power spectra associated with this bispectrum correspond to 
various choices of two fields $\rm X$ and $\rm Y$ to 
correlate with the third field $\rm Z$. We can construct three different power 
spectra related to this given bispectra. 
\ben
&& 
%\myC_l^{S\Theta,E} = \sum_{l_1l_2} 
%J_{l_1l_2l}[b^E_{l_2l_1}]; \qquad
\myC_\ell^{S\Theta,E} = \sum_{\ell_1\ell_2} J_{\ell_1\ell_2\ell} 
[b^E_{\ell_2\ell_1}]; \qquad
%C_l^{S\Theta,E} = \sum_{l_1l_2} 
%J_{l_1l_2l} b^2_{l_2l_1}; \qquad 
\myC_\ell^{SE,\Theta} = \sum_{\ell_1\ell_3} 
J_{\ell_1\ell\ell_3}[b^E_{\ell\ell_1}]; \qquad
\myC_\ell^{E\Theta,S} = \sum_{\ell_2\ell_3} J_{\ell\ell_2\ell_3} 
[b^E_{\ell_2\ell}]; \label{eq:trio}\qquad\\
&& J_{\ell_1\ell_2\ell}=\left ( {\Sigma_{\ell_1}\Sigma_{\ell_2} \over 4 \pi }\right )
\left ( \begin{array}{ c c c }
     \ell_1 & \ell_2 & \ell \\
     0 & 0 & 0
  \end{array} \right )^2.
\label{eq:diff_skew}
\een
Similar results will hold for expressions involving B-type polarisation. The
advantage of using a PCL approach is related to the fact that it does not
depend on detailed modelling of the target theoretical model. It is extremely
fast and is only limited by the speed of harmonic transforms. 
In Fig.~\ref{fig:pcl} we plot the PCL skew-spectra for different reionization history models. Here, we can see more clearly
the dependence of the amplitude of the skew-spectra on the redshift distribution of the tracers discussed in the 
previous section.

\subsection{Defining optimal weights and Near-Optimal Esimators}
The weights required for the construction of an optimal estimator need the theoretical modelling
of the bispectra that is being probed.
\begin{equation}
{S}_\ell^{\rm XY,Z} = \sum_{\ell_1\ell_2} {\hat B}^{\rm XYZ}_{\ell\ell_1\ell_2} B^{\rm XYZ}_{\ell\ell_1\ell_2} {1 \over \myC^{\rm XX}_{\ell}}
{1 \over \myC^{\rm YY}_{\ell_1}}{1 \over \myC^{\rm ZZ}_{\ell_2}}; \qquad {\rm X, Y, Z} = (\Theta, \rm E, S) \quad or \quad 
(\Theta, \rm B, S).
\label{eq:optimum}
\end{equation}
The model bispectrum $B^{\rm XYZ}_{\ell\ell_1\ell_2}$ is a function of reionization history.
The estimator defined above is designed to maximise the power spectra $S_\ell^{\rm XY,Z}$
when data is closer to theoretical expectation. The framework also allows
checks for cross-contribution from different alternative models of reionization and
analysis of the extent to which they can be separated. 

For the construction of these fields we define following set of fields:
\ben
&& {\cal A}_{\ell m}(\oh) = \left [{1\over {\myC_\ell^{\rm TT}} }\right ]\Theta_{\ell m} ; \quad
%{\Theta_{\ell m} \over {\myC_\ell^{\rm TT}} }; \quad
{\cal B}_{\ell m}(\oh) = \left[ {1 \over {\myC_\ell^{\rm EE}}} \right ]; \quad
%{\cal B}_{\ell m}(\oh) = {\rE_{lm} \over {\myC_\ell^{\rm EE}}};
{\cal C}_{\ell m}(\oh,r) = \left [{\Psi_{\ell}(r)\over \myC_\ell^{\rm SS}} \right ]\rS_{\ell m}; \nn \\
%{\cal C}_{\ell m}(\oh,r) = \Psi_{\ell}(r){\rS_{\ell m}\over \myC_\ell^{\rm SS}}; \nn \\
&& \Psi_{\ell}(r)=P_{gs}  \left ({\ell \over d_{A}(r)} \right ) g(r) W(r) {G^2(r) \over d^2_A(r)}I_\ell^E(r).
\label{eq:auxiliary}
\een
\cb{The weights used in constructing our estimator are displayed within square brackets above.} The angular power spectra $\myC_\ell^{\rm TT}$ appears due to approximate inverse
variance weighting. The field ${\cal A}(\oh,r)$ is essentially the reconstituted CMB temperature 
field from the harmonics $\Theta_{\ell m}$, with no radial dependence. Likewise, the field ${\cal B}_{\ell m}(\oh,r)$, which is scaled $E_{\ell m}$ has no radial dependence either and depends only on the E-type polarization power-spectrum  ${C_\ell^{\rm EE}}$.
The third field is related to the tracer field, and by construction has a radial dependence 
through the weight $\Psi_{\ell}(r)$ introduced above. All power-spectra include signal and noise.

For generic fields if we compute the harmonic transform of their
product field $[{\cal A}{\cal B}](\oh,r)$ we can write them in terms of individual harmonics:
\be
[{\cal AB}]_{\ell m}(r) = I_{\ell\ell_1\ell_2}
\left ( \begin{array}{ c c c }
     \ell_1 & \ell_2 & \ell \\
     m_1 & m_2 & m
  \end{array} \right )
\left ( \begin{array}{ c c c }
     \ell_1 & \ell_2 & \ell \\
     0 & 0 & 0
  \end{array} \right ){\cal A}_{\ell m}(r){\cal B}_{\ell m}(r); \qquad I_{\ell_1 \ell_2 \ell} = \sqrt{\Sigma_{\ell_1}\Sigma_{\ell_2}\Sigma_{\ell} \over 4 \pi}.
\ee
We have retained the radial dependence to keep the derivation generic. The required estimator is then constructed by simply cross-correlating it with
the harmonics ${\cal C}_{\ell m}(r)$ of the third field and performing a line of sight integration.
\be
S_\ell^{\cal AB,C}(r) = {1 \over \Sigma_{\ell}} \sum_{m} [{\cal AB}]_{\ell m}(r){\cal C}_{\ell m}(r); \qquad 
S_\ell^{\cal AB,C} = \int dr \; S_\ell^{\cal AB,C}(r).
\label{eq:opt_skew_spec}
\ee
The estimator described above is optimal for all-sky coverage and homogeneous noise, but it needs to be multiplied by a factor $f^{-1}_{\rm sky}$
in case of partial sky coverage to get an unbiased estimator. Here $f_{\rm sky}$ is the fraction of sky covered.
An optimal estimator can be developed by weighting the observed harmonics using 
inverse covariance matrix $C^{-1}_{\ell m,\ell' m'}$ which encodes information about
sky coverage and noise. Finally the estimator will also have to take into account the target
bispectrum which is used for the required matched filtering. The expression quoted
above is for all-sky, homogeneous noise case. In practice spherical symmetry will be broken 
due to either presence of inhomogeneous noise or partial sky coverage \citep{MuHe10}
which will require (linear) terms in addition to the cubic terms. 

The one-point estimators that one can use are simply weighted sums of these skew-spectra
$S^{\cal ABC}= \sum_\ell (2\ell+1) S_\ell^{\cal AB,C}$. It is also possible to use unoptimised one-point estimator
$S_3^{\cal ABC}= \sum_\ell (2\ell+1) C_\ell^{\cal AB,C}$.

\subsection{Optimal skew-spectra, lensing contamination and signal-to-noise}
Gravitational lensing is the primary source of contamination affecting the study of reionization history using 
the mixed bispectrum $B_{\ell_1\ell_2\ell_3}^{\rm E\Theta S}$ since lensing couples to the tracer field. Following \citet{Cooray04}
we have:
\beqa
&& {\cal B}_{\ell_1\ell_2\ell_3}^{\rm E\Theta S} = {1 \over 2} I_{\ell_1\ell_2\ell_3}
\left [ \left ( \begin{array}{ c c c }
     \ell_1 & \ell_2 & \ell_3 \\
     2 & 0 & -2
  \end{array} \right ) F(\ell_1,\ell_2,\ell_3)C_{\ell_2}^{\Theta E}C_{\ell_3}^{\phi S}
+  \left ( \begin{array}{ c c c }
     \ell_1 & \ell_2 & \ell_3 \\
     0 & 0 & 0
  \end{array} \right )
F(\ell_1,\ell_2,\ell_3)C_{\ell_1}^{\Theta E}C_{\ell_3}^{\phi S} \right ]; \\
&& \qquad F(\ell_1,\ell_2,\ell_3) = [\Pi_{\ell_2} + \Pi_{\ell_3} - \Pi_{\ell_1}]; \quad\quad \Pi_{\ell_i}=\ell_i(\ell_i+1).
\label{eq:lensing}
\eeqa
The cross-power spectra $\myC_\ell^{\phi S}$ between lensing potential $\phi$ and the tracer field
$S$ can be written as follows:
\be
\myC_\ell^{\phi S} = \int dr { W^S(r) \over d_A^2(r) } W^{\phi} \left ( k = {\ell \over d_A(r)};r\right );
\qquad W^{\phi}(k,r) = {-3\Omega_{\rm M}} \left ( {H_0 \over k} \right )^2 
{d_A(r_0-r) \over d_A(r) \; d_A(r_0)}; \qquad r_0 = r(z=1000).
\ee
We can define PCL estimators associated with these lensing bispectra. At the level of
optimal estimators it also possible to access amount of cross-contamination from lensing 
in an estimation of secondary non-Gaussianity. To this purpose we define the skew-spectra such as
\be
S_\ell^{\rm E,\Theta S} = {1 \over \Sigma_{\ell}}\sum_{\ell_2\ell_3} {{\cal B}^{\rm E\Theta S}_{\ell\ell_2\ell_3}{\cal B}^{\rm E\Theta S}_{\ell\ell_2\ell_3} \over \myC_\ell^{\Theta\Theta} \myC_{\ell_2}^{\rm EE} \myC_{\ell_3}^{\rm SS}} .
\label{eq:cont}
\ee
Other skew-spectra can also be defined in a similar manner. The ordinary power spectra that appear 
in the denominator include instrumental noise. 

In the left panel of Fig.~\ref{fig:optimum_skew} we plot the {\em optimal} skew-spectrum $S^{\Theta \rm E, X}_{\ell}$ defined in Eq.~(\ref{eq:opt_skew_spec}) for different reionization history models. To compare to the lensing contamination effect we plot in the right panel Fig.~\ref{fig:optimum_skew} the mixed lensing skew-spectrum defined in Eq.~(\ref{eq:cont}) for the ``slow'' model case. Other reionization history models have similar magnitude. Therefore lensing can be ignored for all practical purposes.

As in \citep{Cooray04} we estimate the signal-to-noise of the bispectrum detection as
\be
\left(\frac{S}{N}\right)^2=\cb{f_{\rm sky}}\sum^{\ell_{max}}_{\ell_1\ell_2\ell_3}\frac{(B_{\ell_1\ell_2\ell_3}^{\rm E\Theta S})^2}{C_\ell^{\Theta\Theta} \myC_{\ell_2}^{\rm EE} \myC_{\ell_3}^{\rm SS}}.
\label{eq:s/n}
\ee
\cb{We compute the signal-to-noise for the different reionization models and tracer fields, 
the results are quoted in Table \ref{table:s2n} for an experiment without instrumental noise and detector noise for Planck 
type experiments: ${\myC}_\ell^{\rm XX}=\myC_\ell^{\rm XX}+{N^{2}_{\ell, \rm X}};\,\,\,\,\rm X=\{\Theta,E\}$ with 
$N_\ell$ being the noise power spectrum that depends on the specific choice of channels which are specified by 
beam and noise characteristics (see Eq.\ref{eq:noise} below). We take the fraction
of sky coverage $f_{\rm sky}=0.8$ and sky resolution is fixed at $\ell_{\rm max}=2000$.}
\be
\cb{N_{\ell,\rm X}^2 = \sum_c{1 \over  (\sigma_{c,\rm X}\phi_c)^2} \exp{[-\ell(\ell+1)\phi^2_c/ (8\; \rm log\, 2)]}.}
\label{eq:noise}
\ee
\cb{The beam and noise parameters $\phi_c$ and $\{\sigma_{c,\rm T}, \sigma_{c,\rm E}\}$ used in our calculations for various 
channels are displayed in Table \ref{table:Planck}. }
\begin{table*}
\begin{center}
\caption{\cb{Planck Survey Parameters}}
\begin{tabular}{|c |c|c| c}
\hline
\hline
Frequency (GHz) & 100  &  143 & 217   \\
\hline
\hline
$\phi_c$ (arcmin) & 10.0 & 7.1  &  5.0  \\
\hline
$\sigma_{c,\rm T}(\mu \rm K)$ & 6.8 & 6.0 & 13.1 \\
\hline
$\sigma_{c,\rm E}(\mu \rm K)$  & 10.9 & 11.4 & 26.7 \\
\hline
\hline
\label{table:Planck}
\end{tabular}
\end{center}
\end{table*} 

\section{Discussion}
\label{sec:disc}

\begin{figure}
\begin{center}
\begin{tabular}{cc}
{\epsfxsize=7 cm \epsfysize=7 cm  {\epsfbox[27 150 584 715]{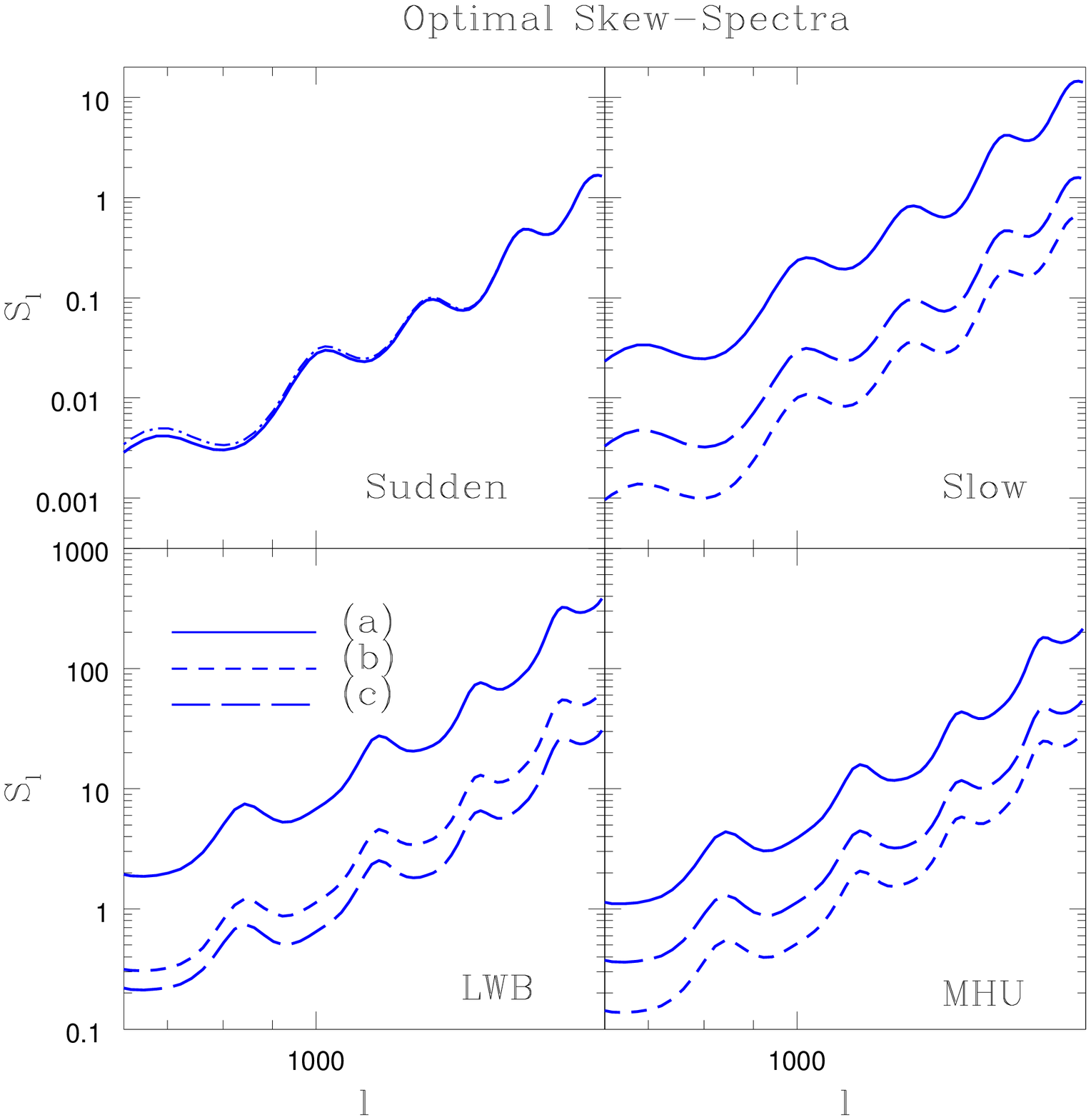}}} &
{\epsfxsize=7 cm \epsfysize=7 cm {\epsfbox[32 436 306 709]{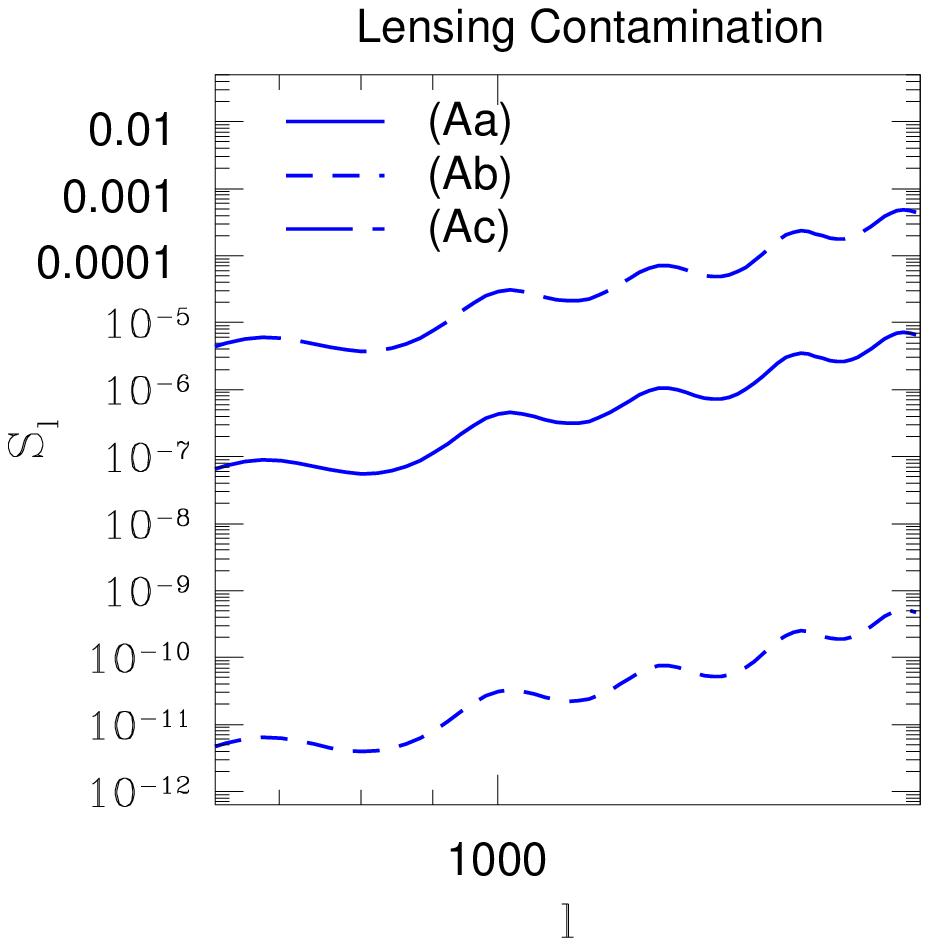}}}
\end{tabular}
\caption{Left panel: optimum skew-spectrum, $S^{\Theta \rm E, X}_{\ell}$, defined in Eq.(\ref{eq:opt_skew_spec}) for different reionization history models and tracer fields. The solid lines corresponds to model (a), short-dashed lines to model (b) and long-dashed lines to model (c). Right panel: lensing contamination to the optimum estimator evaluated using Eq.~(\ref{eq:cont}) in the case of the ``slow'' reionization history model, other scenarios have similar magnitudes.}
\label{fig:optimum_skew}
\end{center}
\end{figure}
Modelling of reionization history is important not only for astrophysical understanding of
the process, but it is also crucial for accurate estimation of cosmological parameters, as
inaccuracy can translate into strongly biased parameters. We have used the mixed bispectrum
involving temperature, polarization and external tracers to map out the reionization history
of the Universe.

{\bf Models and Tracers:} We have used four different models of reionization (A) {\em slow},  (B) {\em sudden}, (C) LWB, (D) MHU to 
study non-Gaussianity induced by reionization. The ionization fraction corresponding to these scenarios are 
plotted in Fig. \ref{fig:models} (left-panel) and Fig. \ref{fig:models2}.
The parametrization of the ionization fractions ${x}_e(r)$ is based on an average redshift of reionization $\bar z$ 
and a transition width $\sigma_z$. Note that we have applied our analysis not only to a single stage reionization, but also to more
complicated scenarios involving multiple stages of reionization.
For {\em external} tracer fields we include three different source populations: (a) proto-galaxy distribution, (b) 21cm-emitting
neutral hydrogen sources, and (c) quasar distribution. We model
the distribution of these sources with redshift (in 3D) using a simple parametric fit. The redshift dependence
of these tracers is plotted in Fig. \ref{fig:models} (right-panel). 

{\bf Estimators, Optimum and Sub-optimum:} The cross-correlation
of these 3D tracer fields with 2D temperature $\Theta$ and E-type polarization maps $\rE$ produces 
a 3D skew-spectrum $S_\ell(r)$ that is presented in Eq. (\ref{eq:opt_skew_spec}). To construct the 3D skew-spectrum we construct the 
harmonics ${\cal A}_{\ell m}(r), {\cal B}_{\ell m}(r)$ and ${\cal C}_{\ell m}(r)$ from $\Theta_{\ell m}$, $\rE_{\ell m}$ and 
$\rS_{\ell m}$ with suitable weighting factors. The weights depends on the background cosmology and the spectrum of fields and 
relevant power-spectra $\myC_{\ell}^{\Theta \rm E}$, $\myC^{\rm EE}_{\ell}$ and $\myC^{\Theta \rm S}_{\ell}$. The exact expressions 
are given in Eq. (\ref{eq:auxiliary}). The power spectra $\myC_{\ell}^{\rm EE}$ and $\myC_{\ell}^{\Theta E}$ are presented in
Fig. \ref{fig:te}.  A line of sight integration is finally used to compute the projected 2D estimator $S_{\ell}$
from its 3D analogue $S_{\ell}(r)$. The other power spectra that are used are plotted in Fig. \ref{fig:te} and \ref{fig:te2}.
The optimal estimator introduced in Eq.(\ref{eq:optimum})  generalize a direct PCL based estimator in Eq. (\ref{eq:direct}). As shown in Eq.(\ref{eq:trio})
depending on the indices that are summed over, we can in principle construct three different 
similar power-spectra. 
%The three maps used in construction are the temperature $\Theta$, electric type polarization $\rm E$
%and a tracer field $\rm S$. 
The three different estimators will carry complementary information and when collapsed to
a one-point estimator (skewness) they can provide an important cross-check for systematics. 
In Eq. (\ref{eq:opt_skew_spec}) we have used estimators that are optimal. Implementation of such 
estimator for inhomogeneous noise and partial sky coverage has already been 
worked out in detail and was adopted for analysis of data from Planck \citep{MuHe10}, mainly
in search for primordial non-Gaussianity.

The results from the computation of the optimum skew-spectra $S^{\Theta \rm E, X}_\ell$ are presented in Fig. \ref{fig:optimum_skew}. The amplitude of 
the skew-spectrum correlates strongly with the redshift range covered by the
tracers.  

{\bf Scatter, Signal-to-Noise and Contamination:} The signal-to-noise of detection for individual models are shown in Table \ref{table:s2n}. For some models it can actually reach  $({\rm S/\rm N}) \approx 70$ but typically
remains at a level of $({\rm S/\rm N}) \approx 20$, thus indicating that such measurements can distinguish among different reionization scenarios at high statistical significance. In Fig. \ref{fig:pcl} we 
present the optimal skew-spectra results for MHU and LWB models. 
\begin{table*}
\begin{center}
\caption{The signal to noise $({\rm S/N})$ of the optimum skew-spectrum estimator for four different
reionization scenarios up to $\ell_{\rm max}=2000$ \cb{(without detector noise)} are shown for various tracer fields. In our estimation we assume all-sky coverage and PLANCK errors. In the top row, results for two different scenarios (A) and (B), which correspond to ``slow'' and ``sudden'' transitions are shown. The bottom row displays results for LWB and MHU reionization. \cb{The numbers in the parentheses correspond to the
realistic values of $({\rm S/N})$ with Planck type noise and sky coverage $f_{\rm sky}=0.8$.}}
\begin{tabular}{|c |c|c| c}
\hline
\hline
& a & b & c \\
\hline
A (slow)&  68.6 \cb{(36.5)} &  14.5 \cb{(7.5)} & 23.0 \cb{(13.0)}   \\
\hline
B (sudden) & 23.4 \cb{(12.6)} & - &  23.5 \cb{(13.2)} \\
\hline
LWB & 76.6 \cb{(39.0)} & 31.5 \cb{(15.8)} & 22.0 \cb{(12.2)} \\
\hline
MHU  & 57.6 \cb{(29.1)} & 21.2 \cb{(10.5)} & 29.5 \cb{(15.7)} \\
\hline
\label{table:s2n}
\end{tabular}
\end{center}
\end{table*} 
The main source of contamination to our estimator is from lensing of CMB. To compute the level of contamination
we have used the ``slow'' model for various tracer fields. We find contamination to be several order
of magnitude lower than the signal. For other models we expect similar result.

\section{Conclusion}
\label{sec:conclu}
The free electron population during reionization epoch re-scatters the local
CMB temperature quadrupole and generates an additional polarization signal at small angular (arcminutes) scale.
Due to their small amplitude, this contribution cannot be studied using the CMB temperature-polarization
cross-spectra. However, additional information regarding the temporal evolution of the spatial variation
of the free electron density can be gained by studying the three-way correlation between temperature anisotropy, polarisation
and an external field, which can act as a tracer field for the free electron density. In harmonic space the
associated mixed bispectrum can be used to constrain models of reionization.
Estimation of individual modes of bispectra are dominated by noise, so the majority of studies in the past have used the skewness, which compresses all available modes to a single high (S/N) number, but this may mask the reionization history. Here we have shown how the recently proposed skew-spectra can be used to discriminate between models of reionization.
We find that the amplitude of the skew-spectra correlates strongly with the epoch
of reionization as well as the redshift distribution of tracers.
We have studied four different models of reionization and
three realistic tracers. We find that the use of multiple tracers can be very powerful in probing the
redshift evolution of the ionization fraction. Most of the signal comes from high $\ell$
hence surveys and tracers with limited sky coverage can also provide valuable information
and all-sky coverage is not an absolute necessity. Our results correspond to a Planck-type
beam but experiments with even higher angular resolution will be able to achieve
higher (S/N). In principle with judicious choice of different tracers it will be
possible to map out the entire ionization history. 

We develop both the direct or PCL-based estimators as well as inverse covariance-weighted
optimal estimators. For each choice of tracer field, we develop three set of estimators
for cross-validation in Eq. (\ref{eq:diff_skew}). \cb{The contamination from weak-lensing was found to be negligible. In case
of an ideal experiment without detector noise, depending on
the redshift distribution of the tracer field, the (S/N) for detection of skew-spectra 
can reach relatively high values in most scenarios typically ${\cal O}(20)$, and even higher for some scenarios.
For Planck type experiment the (S/N) for most scenarios is typically ${\cal O}(10)$.}
 
In the text of the paper we have used the visibility function as our primary variable 
to describe the reionization history.  In the Appendix we detail equivalent results
for optical depth instead .The patchy reionization induces non-Gaussianity
both due to patchy screening as well as Thomson scattering. 
We show that the estimators for reconstructing fluctuations in optical depth
can be cross-correlated with external data sets, and the resulting estimators
are similar to the skew-spectra but with different weights.
These estimators that work with minimum variance reconstruction of optical depth
are however are not optimal and differ from the
corresponding PCL estimators. Cross-correlating with tracers which have redshift 
information has the advantage of distinguishing different histories of
reionization. It is generally believed that the polarization from late-time reionization 
by patchy screening is small compared with polarization due to late-time
Thomson scattering during reionization. However recent studies based on power-spectrum analysis 
have shown that at small angular scales both effects are comparable \citep{DHS09}.
We derive the bispectrum generated by patchy screening of primary as well
as by late-time Thomson scattering. 

The primary motivation of this paper was to devise a method to distinguish between different reionization
histories of the Universe, using non-Gaussianity induced by the fluctuations in optical depth.   This has been achieved, 
using mixed bispectra of CMB temperature and polarisation fields, along with one or more foreground tracers of 
free electron density.  Using skew-spectra, originally devised for studies of primordial non-Gaussianity, we find that different reionization history models 
can be distinguished with this method with high signal-to-noise.

\cb{We have assumed a perfect subtraction of all foregrounds to arrive at our results. 
Needless to say, that, as in any study using CMB data, unsubtracted residuals from the component 
separation step of the data reduction pipeline can seriously bias conclusion drawn 
using techniques presented here.}

%The primary motivation of this paper was to devise a method to distinguish between different reionization
%histories of the Universe, using non-Gaussianity induced by the fluctuations
%in optical depth. The estimator for optical depth fluctuations is 
%mathematically similar to the fluctuations in the gravitational lensing potential.
%We show how to cross-correlate the minimum variance estimator of the
%potential fluctuation with an external tracer with redshift information.
%The resulting cross-spectra are shown to be equivalent to a skew-spectrum
%for the lensing induced non-Gaussianity.

%
\section{Acknowledgements}
\label{acknow} 
%The initial phase of this work was completed when DM
%was supported by a STFC rolling grant at the Royal Observatory,
%Institute for Astronomy, Edinburgh. DM also acknowledges support
%from STFC standard grant ST/G002231/1 at School of Physics and
%Astronomy at Cardiff University where majority of this work was performed.
\cb{DM acknowledges support from the
Science and Technology Facilities Council (grant numbers ST/L000652/1).}
We would like to thank Erminia Calabrese, Patrick Valageas, Asantha Cooray, Antony Lewis and Joseph Smidt
for useful discussions. The Dark Cosmology Centre is funded by the Danish National Research Foundation.
P.S.C. is supported by ERC Grant Agreement No. 279954.
\bibliography{paper.bbl}
\appendix
\section{Bispectrum from patchy Reionization}
\label{sec:patch}
In this appendix we show how the reconstruction of the optical depth $\tau$ and lensing potential $\phi$
studied in the literature \citep{DS09,DHS09,Meerburg13} using quadratic potential is linked to our PCL estimator and the optimum estimator used in the text of the paper. We will consider contribution from patchy screening and lensing of the CMB. See e.g. \cite{Weller99, L01, Dore07} for various aspects of patchy reionization.
\subsection{Patchy Reionization}
Reionization can introduce {\em screening} of the temperature and polarization from the surface of last scattering \citep{DS09}.
Individual line of sight temperature and polarization gets multiplied by $e^{-\tau(\oh)}$ where $\tau$
is the optical depth towards the direction $\oh$. In case of inhomogeneous reionization (IR) this effect
is known to generate addition B-mode polarization \citep{{DHS09}}. In the text of the paper the
visibility function $g(r)$ was treated primary variable. Equivalently, following \cite{DS09} the optical depth $\tau$ will be considered below as the primary variable instead.
The optical depth to a radial comoving distance $r$ along the line of sight is given by:
\ben
%&& \tau(\oh,r) = \sigma_{\rm T}\,n_p\,\int_0^r {dr \over a^2} x_{\rm e}(\oh,r);\\
&&\tau(\oh,z) = \sigma_{\rm T} n_{{\rm p},0} \int_0^z dz'\; {(1+z')^2 \over H(z')}\; x_{\rm e}(\oh,z').
\label{eq:def}
\een
Here $n_{{\rm p},0}$ is the number density of protons at redshift $z=0$ and $x_{\rm e}(\oh,z)$ is the ionization fraction.
\ben
&& p_{\pm}(\oh)= ({ q} \pm i{ u})(\oh) = \int_0^{\infty} \, dr \, \dot \tau \, \exp[-\tau(\oh,r)] \, S^{\pm}_{\rm pol}(\oh,r); \quad\quad
S^{\pm}_{\rm pol}(\oh,r) = - {\sqrt{24\pi} \over 10} \sum_{m=-2}^2 {}_{\pm 2}Y_{2m}(\oh,r) \Theta^{}_{2m}.
\een

Writing $x_{\rm e}(\oh,r)$ as a sum of redshift dependent angular average $\bar x_{\rm e}(r)$ and a fluctuating term
$\delta x_{\rm e}(\oh,r)$ i.e. $x_{\rm e}(\oh,r) = \bar x_{\rm e}(r) + \Delta x_{\rm e}(\oh,r)$.
Expanding $p_{\pm}(\oh)$ in a functional Taylor series we can write:
%\ben
%&& p_{\pm}(\oh) = p_{\pm}(\oh) + \sigma_T n_e \int 
%\een
\ben
&& p_{\pm}(\oh) = p_{\pm}^{(0)}(\oh) + \sigma_{\rm T} n_e \int {dr \over a^2} \Delta x_{\rm e} \;
p^{(1)}_{\pm}(\oh,r).
\een
The zeroth-order term $p_{\pm}^{(0)}(\oh)$ is the polarization from recombination and homogeneous 
reionization. The first-order correction  $p_{\pm}^{(1)}(\oh)$ is due to inhomogeneous reionization:
\ben
&& p^{(1)}_{\pm}(\oh,r) = \int_r^\infty dr' {\delta(P_{\pm}(\oh))\over \delta \tau(r')}
= \left [e^{-\tau} S_{\pm}(\oh,r) -\int_r^{\infty} dr' \dot\tau\; e^{-\tau}S(\oh,r') \right ].
\label{eq:hom_inhom}
\een
The two terms correspond to \underline{\em screening} and Thomson \underline{\em scattering} respectively. For temperature and polarization
we have:
\ben
p_{\pm}(\oh) = p_{\pm}^{(0)}(\oh) +\sum_i \delta\tau^{(i)}(\oh)\; [p^{(1)}_{\pm}(\oh)]^{(i)}; \quad\quad
%\een
%Similar calculation for temperaure is more complicated but follows very similar steps.
%\ben
\Theta(\oh) = \Theta_0(\oh) + \sum_i \delta\tau^{(i)}(\oh) \Theta^{(i)}_1(\oh).
\een
The discrete version of Eq.(\ref{eq:hom_inhom}) that involves redshift binning can be introduced by restricting the integral in Eq.(\ref{eq:def}) into a particular redshift interval. The contributions from a particular tomographic bin is denoted as $\tau^{(i)}$, $\Theta^{(1)}$ and  
$[p^{(1)}_{\pm}]^{(i)}$. 
%$\delta \tau^{(i)}(\oh) = \sigma_{\rm T} n_e \int_{r^{(i)}_{\rm min}}^{r^{(i)}_{\rm max}} \; dr\; x_{\rm e}(\oh,r)/a^2(r)$.
To define an estimator for $\tau$ we can write for arbitrary fields $\rm X$ and $\rm Y$:
\ben
&& \langle {\rm X}_{\ell_1 m_1}{\rm Y}_{\ell_2 m_2} \rangle = (-1)^{m_2}\myC_{\ell_1}^{\rm X Y} \delta_{\ell_1\ell_2}\delta_{m_1 -m_2}
+ \sum_{\ell_1\ell_2}\Gamma^{\rm XY}_{\ell_1\ell_2\ell}\left ( \begin{array}{ c c c }
     \ell_1 & \ell_2 & \ell \\
     m_1 & m_2 & m
  \end{array} \right ){\Delta \tau}_{\ell m}; \quad ({\rm X,Y}) \in {\Theta,\rm E,B};\\
&& \hat\tau_{\ell m} = {\rm N}_{\ell} \sum_{\ell_1 m_1}\sum_{\ell_2 m_2} \Gamma^{\rm X Y}_{\ell_1\ell_2\ell} \left ( \begin{array}{ c c c }
     \ell_1 & \ell_2 & \ell \\
     m_1 & m_2 & m
  \end{array} \right ) {{\rm X}_{\ell_1m_1} \over \myC^{\rm XX}_{\ell_1}} {{\rm Y}_{\ell_2m_2} \over \myC^{\rm YY}_{\ell_2}}.
\een
The power spectra $\myC^{\rm XX}_{\ell}=\langle X_{\ell m}X^*_{\ell m}\rangle$ and $\myC^{\rm YY}_{\ell}=\langle Y_{\ell m} Y^*_{\ell m}\rangle$
also include respective noise. The normalisation ${\rm N}_{\ell}$ is fixed by demanding that the estimator be unbiased. 
The mode coupling matrix $\Gamma$ for various choices of variables $\rm X$ and $\rm Y$ are listed below \citep{DS09}:
\ben
&& \Gamma^{\rm \Theta\Theta}_{\ell_1\ell_2\ell} = 
\left [ \myC^{\rm \Theta_0\Theta_1}_{\ell_1} + \myC_{\ell_2}^{\rm \Theta_0 \Theta_1} \right ] R^{000}_{\ell_1\ell_2\ell}; \\
&& \Gamma^{\rm EE}_{\ell_1\ell_2\ell} = 
\left [ \myC^{\rm E_0E_1}_{\ell_1} + \myC_{\ell_2}^{\rm E_0 E_1} \right ] 
\left [ R^{-220}_{\ell_1\ell_2\ell} +  R^{2-20}_{\ell_1\ell_2\ell} \right ]; \\
&& \Gamma^{\rm \Theta E}_{\ell_1\ell_2\ell} =  {\myC_{\ell_1}^{\Theta_0 E_1} \over 2}\left [ R^{-220}_{\ell_1\ell_2\ell} +  R^{2-20}_{\ell_1\ell_2\ell} \right ] + \myC^{\Theta_1E_0}_{\ell_2} R^{000}_{\ell_1\ell_2\ell}; \\
&& \Gamma^{\rm \Theta B}_{\ell_1\ell_2\ell} =  {\myC_{\ell_1}^{\Theta_0 E_1} \over 2i}\left [ R^{-220}_{\ell_1\ell_2\ell} -  R^{2-20}_{\ell_1\ell_2\ell} \right ] + \myC^{T_1E_0}_{\ell_2} R^{000}_{\ell_1\ell_2\ell}; \\
&& \Gamma^{\rm EB}_{\ell_1\ell_2\ell} = {\myC_{\ell_1}^{\Theta_0 E_1} \over 2i}\left [ R^{-220}_{\ell_1\ell_2\ell} -  R^{2-20}_{\ell_1\ell_2\ell} \right ];\\
&& R^{s_1s_2s_3}_{\ell_1\ell_2\ell} = \sqrt{\Sigma_{\ell_1}\Sigma_{\ell_2}\Sigma_{\ell_3} \over 4\pi}
\left ( \begin{array}{ c c c }
     \ell_1 & \ell_2 & \ell_3 \\
     s_1 & s_2 & s_3
  \end{array} \right ).
\een
In general a minimum variance estimator for $\tau$ can be obtained by including temperature and polarization
maps and is expressed as:
\ben
\hat \tau_{\ell m} = {{\rm N}_{\ell} \over 2 }\sum_{\rm X Y} \sum_{\rm X' Y'}\sum_{\ell_1m_1}\sum_{\ell_2m_2} \Gamma^{\rm X Y}_{\ell_1\ell_2\ell}
\left ( \begin{array}{ c c c }
     \ell_1 & \ell_2 & \ell \\
     m_1 & m_2 & m
  \end{array} \right ) [C^{\rm XX'}]^{-1}_{\ell_1m_1} {\rm X}'_{\ell_1m_1,\ell_1'm_1'} 
[C^{\rm YY'}]^{-1}_{\ell_2m_2,\ell_2'm_2'} {\rm Y}'_{\ell_2'm_2'}.
\label{eq:quad}
\een

We an cross-correlate the above minimum variance reconstruction of $\tau$ with an external tracer $\rm Z$ with redshift information:
\ben
\myC_{\ell}^{\tau \rm Z} = {1 \over 2\ell +1} \sum_{\ell_1\ell_2} \Gamma^{\rm XY}_{\ell_1\ell_2\ell}\;
{B_{\ell_1\ell_2\ell}^{\rm XYZ} \over \myC^{\rm XX}_{\ell_1}\myC^{\rm YY}_{\ell_2}}.
\label{eq:recon}
\een
The resulting estimator is sub-optimal and similar to our 
PCL defined in Eq.(\ref{eq:direct}) and optimal estimator defined in Eq.(\ref{eq:diff_skew}).
A specific model of reionization and its cross-correlation with an external tracer is required for 
explicit computation of $\myC_{\ell}^{\tau \rm Z}$.

Some of the results presented here will be useful in probing reionization
using the kinetic Sunyaev-Zeldovich effect (\cite{Park13} and references therein).
\subsection{Patchy Screening Induced non-Gaussianity and resulting non-Gaussianity}
As discussed above, the scattering of CMB during reionization out of the line-of-sight suppresses the primary temperature polarization
anisotropy from recombination as $\exp(-\tau)$ where $\tau$ is the Thomson optical depth. If $\tau$ varies across
the line of sight this suppression itself induces anisotropy in temperature $\Theta(\oh)$
and polarization Stokes parameters. Following \cite{DHS09} the amplitude modulation due to patchy screening an be expressed as:
\ben
&& \Theta(\oh) = \exp[-\tau(\oh)]\Theta^{(\rm rec)}(\oh); \quad\quad
(\rm Q\pm iU)(\oh) = \exp[-\tau(\oh)] (\rm Q\pm iU)^{(\rm rec)}(\oh)
\een
where
\ben
&& {\rm \Theta}^{(\rm rec)}(\oh) = \sum_{\ell m} \Theta^{\rm (rec)}_{\ell m} Y_{\ell m}(\oh); \quad\quad
(\rm Q\pm i \rm U)^{(\rm rec)}(\oh) = -\sum_{\ell m}(E^{\rm (rec)}_{\ell m} \pm i\rB^{\rm (rec)}_{\ell m}) [{}_{\pm 2}Y_{\ell m}(\oh)].
\een
If we separate the monopole $\bar \tau$ and fluctuating component of the optical depth $\tau$, we can write
$\tau(\oh) = \bar\tau+ \sum_{\ell m}\tau_{\ell m} Y_{\ell m}(\oh)$.
Assuming $\delta \tau \equiv (\tau(\oh)-\bar\tau) \ll 1$ in the harmonic domain we have
$\Theta_{\ell m} = \exp(-\bar\tau)\Theta^{\rm rec}_{\ell m}+\Theta^{\rm scr}_{\ell m}$ and similarly for
other harmonics $E_{\ell m}$ and $B_{\ell m}$. Here $\Theta^{\rm (rec)}$ is same as $\Theta^{(0)}$ of previous section.
The screening contribution can be expressed as a function
of fluctuation in $\tau$ and respective fields at recombination \citep{DHS09}:
\ben
&& \Theta_{\ell m}^{(\rm scr)} = -\exp(-\bar \tau) \sum_{\ell' m' \ell'' m''}\delta\tau_{\ell'' m''}
\Theta^{\rm (rec)}_{\ell' m'} \; R^{000}_{\ell \ell'\ell''}\left ( \begin{array}{ c c c }
     \ell & \ell' & \ell'' \\
     m & m' & m''
  \end{array} \right ); \\
&& \rE_{\ell m}^{(\rm scr)} = -\exp(-\bar \tau) \sum_{\ell'm'\ell''m''}\delta\tau_{\ell'' m''} \rE^{(rec)}_{\ell'm'}
\epsilon_{\ell\ell'\ell''} \; R^{220}_{\ell \ell'\ell''}\left ( \begin{array}{ c c c }
     \ell & \ell' & \ell'' \\
     m & m' & m''
  \end{array} \right ); \\
&& i\rB_{\ell m}^{(\rm scr)} = -\exp(-\bar \tau) \sum_{\ell'm'\ell''m''}\delta\tau_{\ell'' m''} \rE^{(rec)}_{\ell'm'}
\beta_{\ell\ell'\ell''} \;  R^{220}_{\ell \ell'\ell''}\left ( \begin{array}{ c c c }
     \ell & \ell' & \ell'' \\
     m & m' & m''
  \end{array} \right ); \\
&& \epsilon_{\ell\ell'\ell''} = {1 \over 2}[ 1+ (-1)^{\ell+\ell'+\ell''}]; \quad \beta_{\ell\ell'\ell''} = 
{1 \over 2}[ 1 - (-1)^{\ell+\ell'+\ell''}].
\een
The mixed bispectrum involving $\Theta_{\ell m}$, $\rE_{\ell m}$ and an external tracer field ${\rm X}_{\ell m}$ has a vanishing
contribution at the leading order, if we assume vanishing cross-correlation between CMB fluctuations generated at 
recombination and the local tracer field. However, at next-to-leading order the bispectrum can be computed
using a model for cross-spectra involving $\tau$ and $X$, denoted as $\myC^{\tau X}_{\ell}$. The corresponding bispectrum 
takes the following form:
%Ignoring any correlation of $\delta\tau_{\ell m}$ with $\rE_{\ell m}^{\rm rec}$ and $\rB_{\ell m}^{\rm rec}$ fields 
%generated at recombination the lowest order correction to $\myC^{\rm \Theta E,X}_{\ell}$ we can write :
\ben
B^{\rm \Theta \bf EX}_{\ell\ell'\ell''} = \exp(-2\bar \tau)\myC_{\ell'}^{\rm \Theta E}\myC_{\ell''}^{X\rm \tau} 
\left [\epsilon_{\ell\ell'\ell''}W_{\ell\ell'\ell'}^{220} + W_{\ell\ell'\ell'}^{000} \right ].
%\myC^{\rm \Theta E,X}_{\ell} = \exp(-2\bar\tau)\sum_{\ell'\ell''}\myC^{\rm \Theta E}_{\ell'}\myC^{\tau X}_{\ell''}
%I^{000}_{\ell \ell'\ell''} W^{220}_{\ell \ell'\ell''} \epsilon_{\ell\ell'\ell''}.
\een
Corresponding results for other combinations of harmonics can be derived using similar reasoning.
The related optimum skew-spectra can be constructed using these bispectra in Eq.(\ref{eq:optimum}).
Similarly the resulting PCL skew-spectra can be constructed using Eq.(\ref{eq:direct}). 
%The patchiness of reioniation has been studied before \citep{DHS09} for generation of B-type polarization.
\section{Lensing Reconstruction and Contamination}
\label{sec:lensing_recon}
The reconstruction of the lensing potential of the CMB can be treated in an equivalent manner \citep{DS09}.
Minimum variance quadratic estimators in line with Eq.(\ref{eq:quad}) can be constructed using the following coupling functions:
%In the absence of gravity wave and assuming the lensing is generated by only density perturbation
%following results can be derived:
\ben
&& \Lambda^{\Theta\Theta}_{\ell_1\ell_2\ell} = 
\left [ \myC^{\Theta\Theta}_{\ell_1} F^{0}_{\ell_2\ell_1\ell} + \myC_{\ell_2}^{\Theta\Theta} F^{0}_{\ell_1\ell_2\ell} \right ]; 
\label{eq:lensing_start}\\
&& \Lambda^{\Theta \rm E}_{\ell_1\ell_2\ell} = 
{1\over 2}\myC^{\Theta \rm E}_{\ell_1} 
\left [ F^{-2}_{\ell_2\ell_1\ell} +  F^{2}_{\ell_2\ell_1\ell} \right ] + \myC^{\Theta \rm E}_{\ell_2}  
F^{0}_{\ell_1\ell_2\ell} \;\; ; \\
&& \Lambda^{\rm EE}_{\ell_1\ell_2\ell} =  {\myC_{\ell_1}^{\rm EE} \over 2}\left [ F^{-2}_{\ell_2\ell_1\ell} + 
F^{2}_{\ell_2\ell_1\ell} \right ] + {\myC^{\rm EE}_{\ell_2}\over 2}
\left [ F^{-2}_{\ell_1\ell_2\ell} + F^{2}_{\ell_1\ell_2\ell} \right ]; \\
&& \Lambda^{\Theta \rm E}_{\ell_1\ell_2\ell} =  {\myC_{\ell_1}^{\Theta \rm E} \over 2i}\left [ F^{-2}_{\ell_2\ell_1\ell_3} 
- F^{2}_{\ell_2\ell_1\ell} \right ]; \\
&& \Lambda^{\rm EB}_{\ell_1\ell_2\ell} = 
{\myC_{\ell_1}^{\Theta \rm E} \over 2i}\left [ F^{-2}_{\ell_2\ell_1\ell_3} 
- F^{2}_{\ell_2\ell_1\ell} \right ];\\
&& F^{\pm s}_{\ell_1\ell_2\ell} = \sqrt{\Sigma_{\ell_1}\Sigma_{\ell_2}\Sigma_{\ell_3} \over 4\pi}
\left [ -\Pi_{\ell_1}+\Pi_{\ell_2}+\Pi_{\ell_3}\right ]\left ( \begin{array}{ c c c }
     \ell_1 & \ell_2 & \ell_3 \\
     \mp s & \pm s & 0
  \end{array} \right ).
%&& \epsilon_{\ell\ell'\ell''} = {1 \over 2}[ 1+ (-1)^{\ell+\ell'+\ell''}]; \quad \beta_{\ell\ell'\ell''} = 
%{1 \over 2}[ 1 - (-1)^{\ell+\ell'+\ell''}].
\label{eq:lensing_end}
\een
Using these expressions for $\Lambda^{\rm XY}_{\ell_1\ell_2\ell}$ 
in Eq.(\ref{eq:lensing_start})-Eq.(\ref{eq:lensing_end}) we can construct an estimator
for the lensing potential $\phi$ in line with Eq.(\ref{eq:quad}). 
\ben
&& \hat\phi_{\ell m} = {\rm N}_{\ell} \sum_{\ell_1 m_1}\sum_{\ell_2 m_2} \Lambda^{\rm X Y}_{\ell_1\ell_2\ell} \left ( \begin{array}{ c c c }
     \ell_1 & \ell_2 & \ell \\
     m_1 & m_2 & m
  \end{array} \right ) {{\rm X}_{\ell_1m_1} \over \myC^{\rm XX}_{\ell_1}} {{\rm Y}_{\ell_2m_2} \over \myC^{\rm YY}_{\ell_2}}.
\een
The corresponding estimator for the lensing skew-spectrum $\myC^{\phi\rm Z}_{\ell}$ is
\ben
\myC_{\ell}^{\phi \rm Z} = {1 \over 2\ell +1} \sum_{\ell_1\ell_2} \Lambda^{\rm XY}_{\ell_1\ell_2\ell}\;
{B_{\ell_1\ell_2\ell}^{\rm XYZ} \over \myC^{\rm XX}_{\ell_1}\myC^{\rm YY}_{\ell_2}}.
\label{eq:recon2}
\een
We considered the possibility of cross-correlating
external tracers, but it is also possible to construct $\myC^{\tau\tau}_\ell$ or $\myC^{\phi\phi}_{\ell}$
for internal detection using CMB data alone, using the four-point correlation function or equivalently
the trispectrum in the harmonic domain. 
\end{document}